\documentclass[fleqn,usenatbib]{mnras}

\usepackage{newtxtext,newtxmath}

\usepackage[T1]{fontenc}

\DeclareRobustCommand{\VAN}[3]{#2}
\let\VANthebibliography\thebibliography
\def\thebibliography{\DeclareRobustCommand{\VAN}[3]{##3}\VANthebibliography}

\usepackage{graphicx}	
\usepackage{amsmath}	
\usepackage{float}
\usepackage{natbib}
\usepackage{caption}
\usepackage{gensymb}
\usepackage{bm}
\usepackage[output-decimal-marker={,},exponent-product=\cdot]{siunitx}
\usepackage[utf8]{inputenc}
\usepackage{newtxtext,newtxmath}
\usepackage{hyperref}

\newcommand{\mr}{\mathrm}

\title[AGN Jet simulations with SPH]{Active galactic nuclei jets simulated with smoothed particle hydrodynamics}

\author[F. Huško, C. G. Lacey]{
Filip Huško,$^{1}$\thanks{E-mail: filip.husko@durham.ac.uk} 
Cedric G. Lacey$^{1}$
\\
$^{1}$ Institute for Computational Cosmology, Department of Physics, University of Durham, South Road, Durham, DH1 3LE, UK\\
}

\date{Accepted XXX. Received YYY; in original form ZZZ}

\pubyear{2022}

\begin{document}

\label{firstpage}
\pagerange{\pageref{firstpage}--\pageref{lastpage}}
\maketitle

\begin{abstract}
Simulations of active galactic nuclei (AGN) jets have thus far been performed almost exclusively using grid-based codes. We present the first results from hydrodynamical tests of AGN jets, and their interaction with the intracluster medium (ICM), using smoothed particle hydrodynamics (SPH) as implemented in the SWIFT code. We launch these jets into a constant-density ICM, as well as ones with a power-law density profile. We also vary the jet power, velocity, opening angle and numerical resolution. In all cases we find broad agreement between our jets and theoretical predictions for the lengths of the jets and the lobes they inflate, as well as the radii of the lobes. The jets first evolve ballistically, and then transition to a self-similar phase, during which the lobes expand in a self-similar fashion (keeping a constant shape). In this phase the kinetic and thermal energies in the lobes and in the shocked ICM are constant fractions of the total injected energy. In our standard simulation, two thirds of the initially injected energy is transferred to the ICM by the time the jets are turned off, mainly through a bow shock. Of that, $70\%$ is in kinetic form, indicating that the bow shock does not fully and efficiently thermalise while the jet is active. At resolutions typical of large cosmological simulations ($m_\mr{gas}\approx10^7$ $\mr{M}_\odot$), the shape of the lobes is close to self-similar predictions to an accuracy of $15\%$. This indicates that the basic physics of jet-inflated lobes can be correctly simulated even at such resolutions ($\approx500$ particles per jet).
\end{abstract}






\section{Introduction}

Active galactic nuclei (AGN) feedback due to energy release by accreting supermassive black holes (SMBHs) is an important process that contributes to the evolution of galaxies (\citealt{Bower2006}, \citealt{Croton2006}, \citealt{Henriques2015}, \citealt{Lagos2018}). It is thought to be the cause of quenching of star formation in massive elliptical galaxies (\citealt{Springel2005}, \citealt{Martig2009}, \citealt{Sturm2011}).  AGN feedback can take the form of radiative or wind feedback from quasars (\citealt{Sanders1988}, \citealt{Silk1998}, \citealt{DiMatteo2005}), where $5-40$ per cent of the infalling matter is converted into energy and radiated away (\citealt{NovikovThorne1973}, \citealt{Noble2011}). This radiation interacts with the ambient gas by directly heating it, as well as driving outflows through radiation pressure (\citealt{Feruglio2010}, \citealt{Cicone2014}, \citealt{Carniani2015}).

Observations of galaxy clusters reveal the existence of an additional mode of AGN feedback: SMBHs can launch jets of relativistic particles that may traverse large distances (\citealt{Blandford1979}, \citealt{Urry1995}), in some cases larger than a Mpc (\citealt{Dabhade2020}, \citealt{Mahato2021}, \citealt{Heinz2021}). As they travel, these jets are decelerated by and deposit their energy into the hot halo of gas that surrounds the host galaxy (\citealt{McNamara2007}, \citealt{Fabian2012}, \citealt{McNamara2012}). The jets are visible in radio frequencies due to synchrotron emission (\citealt{Biermann1987}, \citealt{Odea1998}, \citealt{Markoff2001}), as well as indirectly in X-ray emission in the form of cavities in the hot gas halo (\citealt{Birzan2004}, \citealt{McNamara2005}, \citealt{Wise2007}). 

AGN jets  are often invoked in order to explain otherwise puzzling observations of galaxy clusters. The cooling rates of hot gas in the central regions of galaxy clusters, inferred from X-ray observations, are large enough that we would generally expect large star formation rates in such environments; this is typically not observed (\citealt{Edge1991}, \citealt{Fabian1994}, \citealt{McDonald2018}), with some exceptions (\citealt{ODea2009}, \citealt{McDonald2015}). Furthermore, observations of emission lines, which we would expect in the presence of a cooling flow, suggest low cooling rates (\citealt{Peterson2003}, \citealt{Bregman2006}). The observed cooling rates (and/or central X-ray luminosities) are closely correlated to the jet powers (\citealt{Rafferty2006}, \citealt{Nulsen2009}, \citealt{Russell2013}, \citealt{Hlavacek-Larrondo}), implying the existence of a feedback loop in galaxy clusters. Jets have also been observed in systems that are smaller than galaxy clusters, such as galaxy groups (\citealt{Sancisi1987}, \citealt{Baldi2009}, \citealt{Randall2011}, \citealt{Werner2019}, \citealt{Eckert2021}) and remnants of galaxy mergers (\citealt{Heckman2016}, \citealt{Merritt}, \citealt{Ivison2012}, \citealt{Shabala2017}). More surprisingly, they have also been observed in disc galaxies (\citealt{Ledlow2001}, \citealt{Singh2015}, \citealt{Nesvadba2021}, \citealt{Webster2021}) and dwarf galaxies (\citealt{Pakull2010}, \citealt{Mezcua2019}, \citealt{Yang2020}, \citealt{Davis2022}), indicating that their effects may be widespread.

Models of galaxy formation in the form of hydrodynamical simulations have in recent years begun to incorporate jet feedback on a cosmological scale (\citealt{Weinberger2017b}, \citealt{Dave2019}, \citealt{Dubois2021}). However, the numerical resolution that can be achieved in such cosmological simulations is often thought to be insufficient to fully capture the impact of the jets (\citealt{Bourne2017}, \citealt{Weinberger2017}). In particular, low resolution simulations do not display instabilities in the jets, which are likely important for their evolution and energetics (\citealt{Perucho2006}, \citealt{Perucho2010}). Many simulations have been carried out on smaller scales (galaxy cluster or group scales) in order to facilitate our understanding of jet propagation, energy deposition and the overall effect of jets as a mode of feedback (e.g. \citealt{Reynolds2006}, \citealt{Mendygral2012}, \citealt{Meece2017}, \citealt{Yang2019}). Some of these suggest that jet-like feedback may be reliable even at at lower resolutions, with the jet energetics (e.g. how much energy is kinetic vs. thermal, how quickly it is transferred to the hot gas halo) insensitive to resolution (\citealt{Weinberger2017}). 

Kinetic, jet-like feedback is currently employed in the following large hydrodynamical, cosmological simulations: IllustrisTNG (\citealt{Weinberger2017b}), SIMBA (\citealt{Dave2019}) and HorizonAGN (\citealt{Dubois2014}). All of these simulations model jet feedback by increasing the velocity of gas particles or cells close to the SMBH by values of order $10^4$ km/s. This is done in discrete injection events whose frequency is determined by some jet power $P_\mathrm{j}$, which is taken to be proportional to the accretion rate onto the black hole: $P_\mathrm{jet}=\epsilon_\mathrm{j}\dot{M}_\mathrm{BH}c^2$. Here, $\epsilon_\mathrm{j}$ is the jet efficiency, a numerical factor that encapsulates the efficiency of energy extraction from the SMBH. The EAGLE hydrodynamical simulation of galaxy formation currently includes only a thermal mode of AGN feedback, implemented as isotropic heating of gas (\citealt{Schaye2015}), which represents the effects of radiative (quasar) feedback. The inclusion of a jet-like kinetic mode of feedback would likely lead to more realistic galaxies in the simulation. The successors of the EAGLE simulation will utilise the SWIFT code (\citealt{Schaller2016}), and its SPHENIX hydrodynamical implementation based on smoothed particle hydrodynamics (hereafter SPH; \citealt{Borrow2022}). 

Before including a jet feedback mode in a large cosmological simulation, it is important to verify that the hydrodynamical code it uses can realistically simulate the effects of jets on the surrounding gas. This can be done by simulating an individual jet episode with a constant jet power, and comparing the behaviour of such a jet with theoretical predictions (see \citealt{Komissarov1998} as an early example). Performing this kind of test is especially important for cosmological simulations using SPH codes, since there has been virtually no effort to simulate individual AGN jets with SPH. We note that some previous SPH simulations have included AGN jets, but these simulations were more complex (e.g. self-consistent feedback in \citealt{Barai2016} or jets as a feedback mechanism in cosmological simulations as employed by \citealt{Choi2015}), and thus harder to compare with theoretical expectations. In recent years, SPH codes have been upgraded in various ways in order to better deal with problems such as fluid mixing and conduction across shocks, that plague traditional SPH schemes (\citealt{Hopkins2015}, \citealt{Menon2015}, \citealt{Wadsley2017}, \citealt{Rosswog2020}, \citealt{Borrow2022}). This is particularly important for jet simulations, since they involve extreme contrasts in fluid properties.

Theoretical studies of constant-power jets, propagating in a gaseous medium with a power-law density profile, predict that all jets start off with a ballistic phase (\citealt{Falle1991}, \citealt{Kaiser1997}, \citealt{Kaiser2007}). During this phase they easily drill through the ambient medium. After the mass of the swept-up medium exceeds that of the jet material, the jet transitions into a self-similar phase, which should always occur at large enough distances (depending on the jet power and mass flux, as well as the ambient medium density). In this phase, the jet material experiences strong shocks and begins to inflate hot lobes of gas (also referred to as 'cocoons', e.g. \citealt{Komissarov1998}, although we use the former term hereafter for consistency). These lobes then collimate the jets of unshocked material. The jet-inflated lobes in this regime expand in a self-similar fashion (with a constant aspect ratio), hence the name. This phase of jet evolution is especially suitable for hydrodynamical tests of jet behaviour since the dependence of jet and lobe properties (such as length) on time, jet power and background density should be very simple (a power law, as long as the background density profile is also a power law). In this paper, we present results on individual jet episodes simulated with SWIFT, using the SPHENIX hydrodynamics scheme. Our focus is on comparing the properties of these jets with theoretical predictions in the self-similar regime of evolution.

We use a jet power and opening angle, as well as properties of the ambient medium, that are close to observed values so that the properties of our simulated jets can be meaningfully compared to observations. We stress, however, that our main aim in this paper is not to make comparisons with observations, but rather with analytical predictions, for the purpose of validating our numerical scheme. We thus favour simplicity over realism in this work. The numerical resolution we achieve, of order $\sim 1$ kpc (within the jet-inflated lobes), is on a par with many similar simulations of AGN jets that use grid-based codes (e.g. \citealt{Yang2019}, \citealt{Smith2021},\citealt{Wang2022}). 


The jet launching velocity, $v_\mr{j}$, is a very important parameter in our simulations, and it has both a physical and numerical role. On the physical side, the launching velocity determines when the jet reaches the self-similar phase of evolution, as well as whether it is in the relativistic regime. In addition, increasing the velocity leads to higher typical temperatures in the jet lobes, as well as lower densities. This latter fact is due to the kinetic energy per particle scaling as $\propto v_\mr{j}^2$, so the total mass and number of particles in the jets and lobes scale as $1/v_\mr{j}^2$, under the assumption that the energy within the jets and lobes is kept fixed. The same scaling ties into the role of the launching velocity as a numerical parameter. Less massive jets and lobes are represented with a smaller number of particles (given a constant jet power), so they are more poorly resolved. 

We do not simulate relativistic or mildly relativistic jets with $v>0.3c$ representing some of the stronger/younger Fanaroff-Riley (FR) II sources (\citealt{Wardle1997}, \citealt{Jetha2006}, \citealt{ODea2009}, \citealt{Snios2018}), which have been the focus of many recent simulation studies (\citealt{Walg2013}, \citealt{Hardcastle2013}, \citealt{Tchekhovskoy2016}, \citealt{Matsumoto2019}, \citealt{Perucho2022}). This is because such jets would be poorly resolved in our simulations. Due to this restriction, we launch jets with subrelativistic velocities of order $0.1c$, representing FRI sources or FRII sources that have either significantly decelerated or entrained significant amounts of ambient material on a kpc scale (\citealt{Bicknell1995}). According to analytical models of jet and lobe evolution, the properties of jets should be largely insensitive to the choice of launching velocity once they reach the self-similar regime (\citealt{Kaiser2007}). The differences between subrelativistic and relativistic jets, in terms of their properties such as the lengths or shapes of the lobes, are minimal (of order tens of percent), and they arise largely from the different adiabatic indices of the lobe material ($4/3$ versus $5/3$). 

The outline of this paper is as follows. In Section \ref{sec:theory}, we summarise existing theoretical predictions for the evolution of jets and the lobes they inflate in the self-similar regime, while in Section \ref{sec:implementation} we discuss our numerical implementation of the jet launching process, as well as the different simulations we have done. In Section \ref{sec:general_properties} we discuss some general properties of our simulated jets, including their morphology, energetics and how well they compare to predictions of analytical models. We also compare with previous simulations of AGN jets. Section \ref{sec:parameter_study} includes a parameter study, where we compare the properties of jets simulated at different numerical resolutions, and with different jet powers, launching velocities and half-opening angles. In all cases, we compare our simulations with theoretical predictions. We also show some results of AGN jets launched into power-law gaseous atmospheres. In Section \ref{sec:conclusion} we summarise and conclude.




\section{Jet and lobe evolution in the self-similar regime}
\label{sec:theory}

In this work, we compare our simulated jets with theoretical predictions for the self-similar regime of jet lobe evolution (e.g. \citealt{Falle1991}, \citealt{Kaiser1997}, \cite{Komissarov1998}; see also \citealt{Begelman1989} and \citealt{Broomberg2011} for alternative, but similar models). In the self-similar picture, the jet is launched from a conical region defined by a half-opening angle $\theta_\mathrm{j}$. The physical quantities that determine the evolution of the jet and its lobes are: 1) power $P_\mathrm{j}$, 2) launching velocity $v_\mathrm{j}$ (or equivalently, mass flux $Q_\mathrm{j}=2P_\mathrm{j}/v_\mathrm{j}^2$), 3) background density $\rho$ and 4) background pressure $p$ (or equivalently, temperature $T$). Note that in our notation, the jet power and mass flux refer to the total, summed over both jets. 

These quantities can be combined to yield two length scales, $L_1<L_2$. The evolutionary phase of a jet, can be determined by comparing its current length, $L_\mathrm{j}$, with those length scales (\citealt{Komissarov1998}). In the initial phase of ($L_\mathrm{j}\ll L_1$), the mass in the jet is large compared to the ambient medium being swept up by the jet. The jet is denser than the ambient medium and it drills through it without significantly being slowed down, due to its large inertia. The jet head moves with a velocity equal to the launching velocity, $v_\mathrm{j}$, and the jet length is thus given by $L_\mathrm{j}=v_\mathrm{j}t$. The jets have not yet reached the self-similar regime while the above condition is true. We will refer to jets that are in this evolutionary phase as 'ballistic'.

The length scale $L_1$ represents the scale at which the mass of the swept up medium becomes comparable to the mass launched into the jet. It is given by
\begin{equation}
L_1 = \frac{1}{\theta_\mathrm{j}}\sqrt{\frac{2}{\pi\rho}\sqrt{\frac{Q_\mathrm{j}^3}{2P_\mathrm{j}}}} = \frac{2}{\theta_\mathrm{j}}\sqrt{\frac{P_\mr{j}}{\pi\rho v_\mr{j}^3}}.
\label{eq:first_scale}
\end{equation}
The second length scale, $L_2$ represents the scale at which the ambient pressure becomes important. It is given by
\begin{equation}
L_2 = \bigg(\frac{P_\mathrm{j}^2 \rho}{p^3} \bigg)^{1/4}.
\label{eq:second_scale}
\end{equation}
Note that these length-scales represent dimensional combinations, and thus do not necessarily include the correct numerical factors. Furthermore, previous work implies that the transition from one regime to another, which should occur once the jet has reached $L_1$ or $L_2$, can be fairly protracted (\citealt{Komissarov1998}).

The majority of observed FR-II sources are expected to satisfy $L_1\ll L_\mathrm{j}\ll L_2$ (\citealt{Komissarov1998}). In this regime, both the mass flux and the ambient pressure are dynamically unimportant. The jet experiences strong shocks and it is effectively slowed down. The jet head velocity is thus expected to be much smaller than the launching velocity $v_\mathrm{j}$. The jet comes into equilibrium with its own lobe (previously shocked particles) through recollimation (reconfinement) shocks. The jet also launches a bow shock. 

Once the mass flux and external pressure are excluded, one cannot form any length-scale from the remaining dynamical quantities, with the exception of time. As a result, the behaviour of the jet-inflated lobes is expected to be self-similar (\citealt{Sedov1959}), and we thus refer to jets that satisfy $L_1\ll L_\mathrm{j}\ll L_2$ as being in the 'self-similar' regime. 
For the rest of the analysis, we assume that the background medium follows a power law in density:
\begin{equation}
\rho(r)=\rho_0\bigg(\frac{r}{r_0}\bigg)^{-\beta}.
\label{eq:density_power_law}
\end{equation}
Including time as a dynamical quantity, one can compute a length scale of the form
\begin{equation}
L = \bigg(\frac{P_\mathrm{j} t^3}{\rho_0 r_0^\beta} \bigg)^{1/(5-\beta)}.
\label{eq:self_similar_scaling}
\end{equation}
 (\citealt{Falle1991}). The actual length of the jet (and the lobe) may differ from $L$ by some numerical factor, which can depend on the dimensionless parameters that govern the jet evolution (half-opening angle $\theta_\mathrm{j}$ and adiabatic index $\gamma$). The actual jet length can be computed from energy conservation if one assumes self-similarity of the lobes and a particular type of geometry. With a cylindrical geometry, the jet length is given by
\begin{equation}
L_\mathrm{j} = c_1\bigg(\frac{P_\mathrm{j}t^3}{\rho_0 r_0^\beta}\bigg)^{1/(5-\beta)},
\label{eq:self_similar_jet_length}
\end{equation}
where $c_1$ is
\begin{equation}
c_1 = \left\{\frac{A^4}{18\pi} \frac{(\gamma_\mathrm{c}+1)(\gamma_\mathrm{l}-1)(5-\beta)^3}{9[\gamma_\mathrm{l}+(\gamma_\mathrm{l}-1)A^2/2]-4-\beta}\right\}^{1/(5-\beta)}.
\label{eq:numerical_coefficient}
\end{equation}
Here, $A$ is the aspect ratio of the lobe (its length divided by radius, also equal to $1/\theta_\mathrm{j}$ for cylindrical jets), and $\gamma_\mathrm{l}$ and $\gamma_\mathrm{c}$ are the adiabatic indices of the lobe and ambient gas, respectively (\citealt{Kaiser2007}).




\section{Numerical implementation}
\label{sec:implementation}

In this work, we use the open-access\footnote{\href{https://swift.dur.ac.uk/}{https://swift.dur.ac.uk/}} SWIFT hydrodynamics and galaxy formation code (\citealt{Schaller2016}), and the SPHENIX hydrodynamics scheme implemented therein (\citealt{Borrow2022}). SPHENIX is an SPH method (\citealt{Monaghan1992}). It includes artificial viscosity, which is necessary in order to capture shocks since traditional SPH is dissipationless. SPHENIX also includes artificial conductivity, which helps reduce unwanted surface tension otherwise present in SPH simulations (\citealt{Argetz2007}, \citealt{Sijacki2012}, \citealt{Nelson2013}), allowing for mixing between flows that are in pressure equilibrium but contrasting in temperature and/or density. 

Both artificial viscosity and conductivity are crucial in our simulations: artificial viscosity because our jets experience strong shocks (in some cases with a Mach number, hereafter $M$, of $M\approx100$), and artificial conductivity since the jet-inflated lobes feature extreme density and temperature contrasts, but are in approximate pressure equilibrium with their surroundings. An artificial viscosity limiter is included to prevent spurious viscosity in shear flows. An artificial conductivity limiter is also included, to prevent spurious energy transfer in all flows.

In our simulations we do not include radiative cooling, gravity, magnetic fields or cosmic rays, since such additional physics might cause deviations from the simple model of self-similar jet and lobe evolution. We have, however, performed runs with gravity and radiative cooling as a consistency check. We found very small differences compared to purely hydrodynamical jets, which shows that the artificial conduction limiter in SPHENIX prevents spurious radiative losses, even with very poorly resolved jets. We do not include relativistic effects, since we do not include very large velocities where these effects occur (the reasoning for this choice is outlined in Section \ref{sec:list_of_sims}). We therefore set the adiabatic index to $\gamma=5/3$ for all gas in our simulations.


\subsection{Jet launching scheme}

AGN jets in an SPH code can be implemented through velocity kicks of gas particles. Given a jet power of interest, $P_\mathrm{j}$, the time interval at which particles need to be kicked is given by
\begin{equation}
\Delta t = \frac{2\times\frac{1}{2}m_\mathrm{gas}v_\mathrm{j}^2}{P_\mathrm{j}}.
\label{eq:timestep}
\end{equation}
Here, $m_\mathrm{gas}$ is the mass of the gas particles in the simulation $v_\mathrm{j}$ is some arbitrary launching velocity, and the factor of $2$ is present to ensure that two particles are always kicked (in opposite directions, ensuring conservation of momentum). The total number of kicking events can be calculated as 
\begin{equation}
N_\mathrm{j}=\frac{T_\mathrm{j}}{\Delta t} = \frac{T_\mathrm{j}P_\mathrm{j}}{m_\mathrm{gas}v_\mathrm{j}^2},
\label{eq:N_jet}
\end{equation}
where $T_\mathrm{j}$ is the lifetime of the jet episode. The larger $N_\mathrm{j}$, the better the jet will be resolved (as one might expect, and as will be clear from our results). 

$N_\mathrm{j}$ can be increased by decreasing the particle mass or launching velocity, or by increasing the total energy launched into the jet (by increasing either the jet power or jet duration). When attempting to simulate the self-similar regime of jet evolution, one also has to keep in mind that the length scale $L_1$ (Eqn. \ref{eq:first_scale}) needs to be small, while the length scale $L_2$ (Eqn. \ref{eq:second_scale}) conversely needs to be large. These represent additional constraints on the choice of parameters characterizing the jets and the ambient medium.

The most natural implementation of AGN jets in SPH would involve kicking particles from the smoothing kernel of the central black hole. The SWIFT hydrodynamical code, which we utilise in this work, includes black holes so this scheme is easy to implement (see \citealt{Husko2022_spin_driven_jet_feedback}, where we use such an implementation in the context of self-consistent accretion and feedback in idealized galaxy groups and clusters). We have attempted this scheme in the present work, and we find that it works in general. However, at very high resolutions (more than $10^4-10^5$ launching events per few dozen Myrs, the kind of resolution we are interested in when testing jet hydrodynamics), this scheme can become computationally expensive and unreliable. In particular, the black hole requires very small time steps between kicking events, smaller than the typical evolutionary time step of the particles kicked into the jet. This can result in particles being kicked more than once. 


For simplicity, we instead populate the initial conditions with a reservoir of particles that are to be used for jet launching, therefore bypassing any issues that might arise in a setup using a BH. In general, the reservoir we use takes the shape of two cones (defined by some half-opening angle $\theta_\mathrm{j}$), placed along both directions of the $z$ axis, up to some maximal radius (10 kpc in all our simulations). We obtain these cones by creating a uniform cube with a grid of particles, and then cut out the desired cones. These particles are not allowed to interact with any other particles until they have been kicked and have cleared the region associated with the reservoir. The particles are launched progressively from the outside-in, so that they can immediately interact with the ambient medium, instead of traveling through the frozen-in reservoir. The total number of particles in the reservoir exactly matches the number to be launched into the jets we are simulating. The density of this reservoir is $\approx10^{-27}$ $\mathrm{g}\mathrm{cm}^{-3}$, which is $\approx10$ times less than the density of the ambient medium. In the case of ballistic jets with null opening angle (only a single simulation), we instead use a spherical reservoir with a radius of 5 kpc, from which particles are launched parallel to the $z-$axis. We have tested a cylindrical reservoir of similar size, but found the differences to be negligible.

\subsection{Physical setup}
\label{sec:setup}

The structure of realistic gaseous haloes, representing the intracluster medium, can be represented using a density profile that is constant in the centre, and falls off as $r^{-\alpha}$ at large distances, with $\alpha\approx2$ (\citealt{Komatsu2001}, \citealt{Croston2008}), at least out to roughly the virial radius (at larger distances the profile drops more sharply, \citealt{Eckert2012}). Many jet simulation studies incorporate profiles similar to this (e.g. \citealt{English2016}, \citealt{Weinberger2017}). While it may be more realistic to launch jets into such a profile, we choose instead a constant density medium ($\alpha=0$) for most of our simulations. We do this since the jet length should scale as $t^{0.6}$ in such a setup, whereas at large distances in a realistic profile ($\alpha=2$), the jet length scales as $t$. This is due to the jet-inflated lobes not behaving self-similarly for $\alpha=2$, whereas for $\alpha=0$, they are firmly in the self-similar regime, provided an appropriate choice of parameters.

We launch most of our jets into the same background medium, with a constant density $\rho_0$ (the value we choose is discussed in the next subsection). In this case we choose periodic boxes that are slightly longer in each dimension ($\approx20\%$) than the predicted sizes of the jets based on the theory outlined in Section \ref{sec:theory}. We find that this works well in all cases, and the jets do not reach the edges of the box by the end of the simulation.

We have performed a few runs where the ambient medium instead features a power-law density profile, such that $\rho\propto r^{-\alpha}$. We restrict ourselves to $\alpha<2$, since the self-similar solution from the previous section is applicable only in this regime. In these cases we use a \cite{Navarro1996} (NFW) background gravitational potential, and we choose the gas pressure (and therefore the temperature) in such a way that the gaseous halo is held in hydrostatic equilibrium. For this purpose we choose NFW parameters representing a galaxy cluster with a halo mass $M_\mathrm{h}=10^{15}$ $\mathrm{M}_\odot$ at redshift $z=0$, virial radius $R_\mathrm{v}\approx2$ Mpc and concentration parameter $c=4$. Since it is impossible to implement a power-law density profile such that the power law is valid all the way to the centre of the halo, we use a cored $\beta$-profile
\begin{equation}
    \rho(r)=\frac{\rho_{0,\beta}}{[1+(r/r_\mr{c})^2]^{3\beta/2}},
\label{eq:beta_profile}
\end{equation}
where we choose the value of $\beta$ to match our desired value of the slope at large radii, $3\beta=\alpha$. We choose a small core, $r_\mr{c}=10$ kpc, which matches the size of our jet reservoir. The normalisation $\rho_{0,\beta}$ is then calculated so that the total mass of the gaseous halo is $15\%$ of the total halo mass. This choice is not necessary in this application, since we are only interested in how our jets compare with theory (and not how realistic they are), but we make it for simplicity.

\subsection{Simulations}
\label{sec:list_of_sims}

In Table \ref{tab:tab0} we summarise the parameters used for all of our simulations. In the first row we specify the fiducial choice of parameters for our constant-density ambient medium simulations. This choice corresponds to: 1) $m_\mathrm{gas}=1.81\times10^5$ M$_\odot$, 2) $P_\mathrm{j}=10^{46}$ erg/s, 3) $v_\mathrm{j}=15000$ km/s and 4) $\theta_\mathrm{j}=10\degree$. We vary all of these parameters, but we do not vary the ambient density, which we choose to be $\rho_0=1.2\times10^{-26}$ g/cm$^{3}$. We also do not vary the jet duration, which we set to $T_\mathrm{j}=100$ Myr.

Our chosen ambient gas density is the typical central density of a galaxy cluster with a halo mass $M_\mathrm{h}=10^{15}$ M$_\odot$, a virial radius $2$ Mpc, and a baryonic mass ratio $0.15$. The initial temperature of this gas is set to $T=10^{7.2}$ K (and this is also the temperature of the gas kicked into the jets). This value is somewhat low for the cores of realistic clusters, but we choose it to ensure that our jets never reach the regime in which ambient pressure is important (Eqn. \ref{eq:second_scale}). In any case, the aim in this work is not to produce perfectly realistic jets, but rather to check that SWIFT can correctly simulate jets. 

Our fiducial mass resolution is 10 times better than the EAGLE simulation (\citealt{Schaye2015}). At this mass resolution, the typical smoothing length (corresponding to spatial resolution) is $\approx1$ kpc in the ambient medium and $\approx7$ kpc in the jet-inflated lobes, which are about $300$ times less dense than the ambient medium. At our highest resolution level, the typical smoothing lengths are instead $\approx0.3$ and $\approx2$ kpc for the two cases. For comparison, the lobe is roughly $400$ kpc long and $50$ kpc in radius at the end of the simulation (see Fig. \ref{fig:fig5}).

The jet power we use is relatively high compared to previous similar simulations (e.g. \citealt{Weinberger2017}, \citealt{Bourne2017}). However, observations imply that jet episodes with such powers are frequent in the most massive galaxy clusters (\citealt{Kino2005}, \citealt{Hlavacek-Larrondo}). Our choice of half-opening angle is somewhat large compared to most real jets (\citealt{Pushkarev2009}), but we choose such a value to ensure that jets are ballistic for as short a time as possible (Eqn. \ref{eq:first_scale}). In addition, observed subrelativistic jets have similarly large opening angles (\citealt{Pushkarev2017}).

We use a non-relativistic launching velocity of order $10^4$ km/s for a few reasons: 1) SWIFT does not include relativistic effects, 2) launching jets with relativistic velocities leads to only small differences (e.g. \citealt{Komissarov1998}, \citealt{English2016}), 3) using velocities of order $10^5$ km/s or higher would result in poorly resolved jets and 4) velocities of order $10^4$ km/s are typically employed in cosmological simulations that include jets (\citealt{Weinberger2017b}, \citealt{Dave2019}). The fiducial mass resolution we have chosen results in a total of $\approx10^8$ particles in the simulation, while the jet power and launching velocity, in combination with the mass resolution and jet duration, yield $\approx40000$ particles launched per jet. The actual number of particles in the jets and lobes may be larger due to ambient particles being swept up. 

When varying any of the four parameters listed at the beginning of this subsection, we keep other parameters fixed. The variations we have done for our constant-density ambient medium case are given in the second row of Table \ref{tab:tab0}. We have simulated jets with numerical resolutions corresponding to ten times worse than the EAGLE simulation ($m_\mathrm{gas}=1.81\times10^7$ M$_\odot$), down to 3160 times better ($m_\mathrm{gas}=5.73\times10^3$ M$_\odot$), differing by factors of 3.16 (logarithmic interval of 0.5). Our highest resolution simulation has a total of $2.8\times10^9$ particles, and $1.4\times10^6$ particles kicked into each jet. The lowest resolution one has only 450 particles per jet. We vary jet powers between $10^{45}$ erg/s and $10^{47}$ erg/s, launching velocities between $3750$ km/s and $60000$ km/s, and half-opening angles between $0\degree$ and $25\degree$.

Finally, in the case of power-law gaseous atmospheres, we have performed three simulations, with $\alpha=0.5$, $\alpha=1$ and $\alpha=1.5$. The parameters of these simulations are listed in the third and final row of Table \ref{tab:tab0}. These power-law cases required a different set of parameters for two reasons. We found that the length scales $L_1$ and $L_2$ were larger and smaller, respectively, with our fiducial choice of jet-related parameters, than they were in the constant-density case. This means that the jets took a longer time to reach the self-similar phase, and would also take a shorter time to exit the same phase due to the external pressure becoming important. 

Given these restrictions, we chose to modify our fiducial parameters in the following way. We launched the jets with: 1) a jet power of $P_\mathrm{j}=10^{47}$ erg/s in order for the length scale $L_2$ to lie at comfortably large distances compared to the self-similar prediction (see Eqn. \ref{eq:second_scale}, although note that its meaning is somewhat moot in power-law atmospheres), 2) a jet velocity of $v_\mathrm{j}=0.2c=60000$ km/s, bringing $L_1$ down to $L_1=5$ kpc (given the new jet power), 3) a jet duration of $T_\mathrm{j}=40$ Myr to prevent the jets from reaching large distances (wishing to avoid both $L_2$ and the virial radius). With these changes to the physical parameters, the number of particles launched into the jets is a quarter of that in our standard constant-density simulations. For this reason we have decreased the particle masses in the power-law simulations (i.e. increased the resolution) by a factor of four. This ensures that the jets are resolved with the same number of particles.

\captionsetup[table]{skip=0pt} 
\begin{table*}
\begin{center}
\caption{List of all simulations and the parameters we use. In the first row we specify the parameters of our fiducial simulation with a constant-density ambient medium. In the second row we specify the parameters that are varied for this same case. In the final row we specify the parameters of the case with a power-law ambient density profile. The parameters are, in order:  1) $m_\mr{gas}$ - mass resolution, 2) $P_\mr{j}$ - jet power, 3) $v_\mr{j}$ - jet velocity, 4) $\theta_\mr{j}$ - jet half-opening angle, 5) ambient medium density $\rho$ - constant value or power law slope and 6) $T_\mr{j}$ - jet duration. }
\label{tab:tab0}
\end{center}

\begin{tabular*}{0.9\textwidth}{@{\extracolsep{\fill}}cccccc}
  \hline 
   $m_\mr{gas}$ $[\mr{M}_\odot]$ & $P_\mr{j}$ [$\mathrm{erg}\hspace{0.3mm}\mr{s}^{-1}$] & $v_\mr{j}$ [$\mathrm{km}\hspace{0.3mm}\mathrm{s}^{-1}$] & $\theta_\mr{j}$ [$\degree$] & $\rho$ [$\mr{g}\hspace{0.3mm}\mr{cm}^{-3}$] & $T_\mr{j}$ [$\mr{Myr}$]  \\
  \hline 
  $1.81\times10^5$ & $10^{46}$ & $1.5\times10^4$ & $10$ & $1.2\times10^{-26}$ & 100\\
  $5.73\times10^{3}-1.81\times10^7$ & $10^{45}-10^{47}$ & $3.75\times10^3-6\times10^4$ & $0-25$ & $1.2\times10^{-26}$ & 100 \\
  $4.53\times10^4$ & $10^{47}$ & $6\times10^4$ & $10$ & $\propto r^{-\alpha}$ , $\alpha=0.5-1.5$ & 40   \\
  \hline
\end{tabular*}

\end{table*}

\subsection{Definition of jet lobe}

All simulations of jets exhibit the so-called lobe, made up from hot, shocked gas that was previously part of the jet. Jets also invariably launch a bow shock that propagates through the ambient medium. The model of self-similar lobe evolution predicts that their aspect ratio should be constant, and it predicts the same for the ratio of energy in the lobes versus the energy added to the ambient medium (as well as for how much energy is in kinetic and thermal forms). In order to test these predictions, it is important to numerically determine which particles belong to what we might visually call the jet or the jet lobe, with remaining particles classified as making up the ambient medium.

\begin{figure*}
\includegraphics[width=0.99\textwidth, trim = 0 5 0 0]{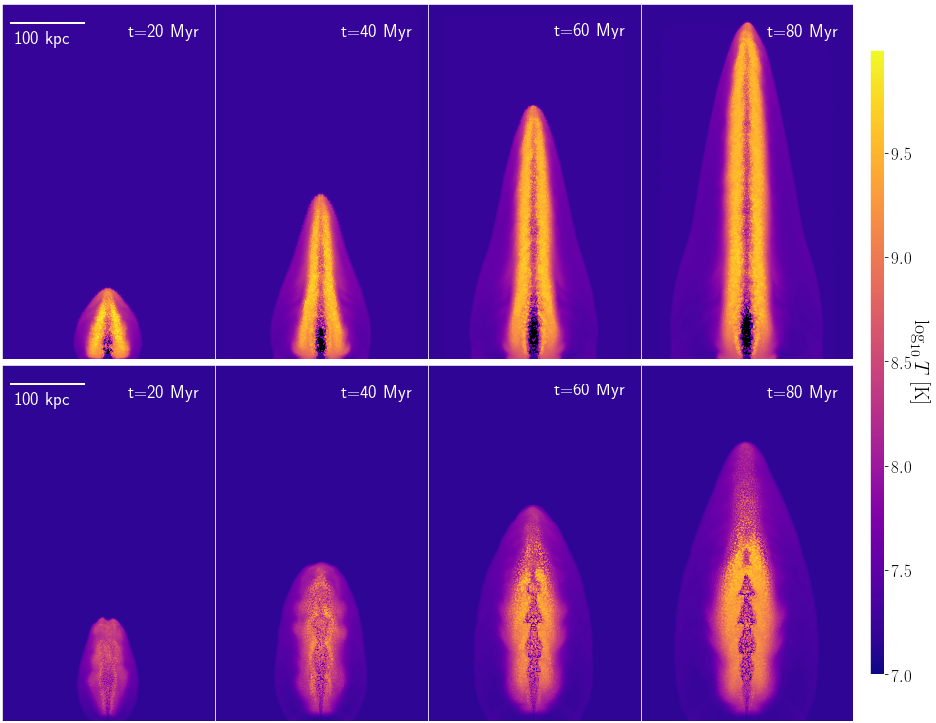}
\caption{Visualisations of the gas temperature distribution for a ballistic jet (effective half-opening angle $\theta_\mathrm{j}=0\degree$, top) and a jet in the self-similar evolutionary phase (half-opening angle $\theta_\mathrm{j}=10\degree$, bottom) at different times. All other parameters correspond to our fiducial choice, given in the first row of Table \ref{tab:tab0}. Colours represent the temperature of the gas, as given by the colour bar. The panels show slices 10 kpc in depth.}
\label{fig:fig1}
\end{figure*}%

Empirically, we find that the peak temperature $T_\mathrm{p}$ achieved by gas particles serves well to define the lobe. We use a threshold value $T_\mathrm{p,min}$, and all particles whose peak temperature was at some point above this value, i.e. $T_\mr{p}>T_\mathrm{p,min}$, are defined as constituting the lobe. This definition is motivated by the fact that particles that reach extremely high temperatures are exclusively located in the lobe, whereas the shocked ambient medium experiences temperature jumps of a factor of several at most.

The appropriate value of $T_\mathrm{p,min}$ varies from simulation to simulation, but it can easily be estimated (to better than an order of magnitude) by assuming that all of the kinetic energy launched into the gas becomes thermalised. From this condition, the characteristic temperature of the lobes is $T_\mr{lobe}\approx\mu m_\mathrm{p}v_\mathrm{j}^2/3k_\mathrm{B}$, with $\mu=0.6$ the mean molecular weight, $m_\mathrm{p}$ the proton mass and $k_\mathrm{B}$ the Boltzmann constant. In reality the lobes are somewhat less hot since not all of the kinetic energy is thermalised. The appropriate value of $T_\mathrm{p,min}$ can thus be expected to be a factor of several times below $T_\mr{lobe}$. We find the particular value that we use by plotting the total mass in the lobe $M_\mathrm{lobe}$ versus $T_\mathrm{p,min}$, the defining peak temperature that determines how many particles will be assigned to the lobe. We find that this dependence exhibits a change in slope at some critical value of $T_\mathrm{p,min}$. Using a larger than the critical one results in fewer and fewer particles (that are part of the lobe) being assigned to the lobe, whereas using a lower one causes ambient medium gas to be assigned to the lobe (usually ambient gas particles near the jet head, where strong shocks are occurring). For our fiducial simulation, with a jet launching velocity of $v_\mr{j}=15000$ km/s, we use $T_\mr{p,min}=5\times10^8$ K. This value is $\approx30$ times larger than the initial temperature of the ambient medium, and $\approx10$ times smaller than the characteristic temperature of the lobe, which is in this case $T_\mr{lobe}\approx5\times10^9$ K.

We also define unshocked, fast-moving, recently launched particles as constituting the lobe, which means that we include the jets into the lobes (for simplicity). Given a launching velocity $v_\mathrm{j}$, we thus define all particles with $\vert \mathbf{v}\vert>0.5v_\mathrm{j}$ as also being part of the lobe. Note that the factor $0.5$ here is fairly unimportant; using any critical velocity value that is significantly above the sound speed of the ambient gas leads to almost all of the unshocked particles being included in the lobes, and none of the ambient medium. Particles not belonging to the jets or the lobes are classified as part of the ambient medium.

\subsection{Measuring the energetics and jet/lobe lengths and radii}
\label{sec:energetics_lengths}

Given the above definition of the lobe, measuring the thermal and kinetic energy gains of both the lobes and the ambient medium is trivial. We calculate the kinetic energy gain of either component by calculating the total kinetic energy in the particles (all of the gas in the simulation is initially not moving), while the thermal energy gain is calculated relative to the initial temperature of all of the gas, which is $T_0=10^{7.2}$ K. 

The length of the lobe is calculated by taking the mean distance from the origin of $n$ farthest particles (ordered along the axis of launching), with $n$ determined to be $3\%$ of all of the particles launched into the jet. With this definition, $n$ is both resolution and time-dependent. For the radius of the lobes, the procedure is similar, but we use cylindrical distances from the launching axis. Note that this choice yields the maximal radius of the lobe, not the average radius. For both the length and the radius of the lobe we use a mean of the values calculated for both of the jets.

\section{Results: general properties of simulated jets}
\label{sec:general_properties}
\subsection{Ballistic jets}

In this section we will discuss properties of jets launched into a constant-density ambient medium. Before focusing on self-similar jet and lobe evolution, it is worth addressing some properties of ballistic jets simulated with SWIFT. According to Eqn. (\ref{eq:first_scale}), jets should remain ballistic at arbitrary distances in the $\theta_\mathrm{j}=0\degree$ case. The top panels of Fig. \ref{fig:fig1} show visualisations of jets from a simulation with a null half-opening angle, simulated at a mass resolution 10 times better than the EAGLE simulation ($m_\mathrm{gas}=1.81\times10^5$ M$_\odot$). The jet power is $P_\mathrm{j}=10^{46}$ erg/s, and the launching velocity $v_\mathrm{j}=15000$ km/s (i.e. these parameters, other than the opening angle, match our fiducial choice listed in Table \ref{tab:tab0}). 

\begin{figure*}
\includegraphics[width=0.99\textwidth, trim = 0 20 0 0]{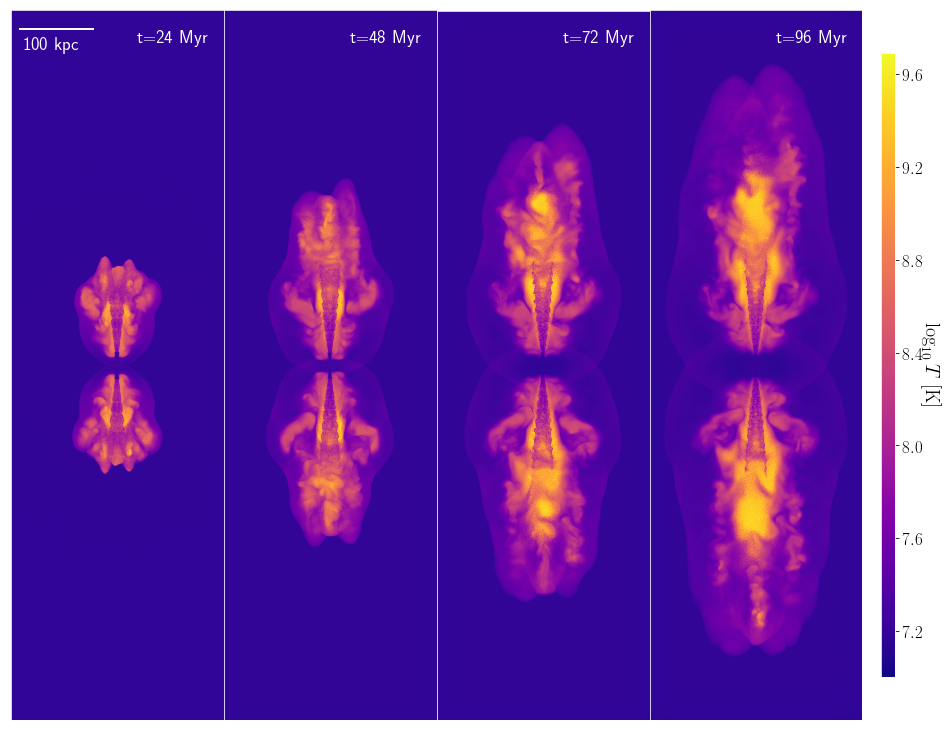}
\caption{Visualisations of the gas temperature distribution in our highest resolution jet simulation, with particle mass $m_\mathrm{gas}=5.73\times 10^3$ M$_\odot$, at different times. All other parameters correspond to our fiducial choice, given in the first row of Table \ref{tab:tab0}. Colours represent the temperature of the gas, as given by the colour bar. The panels are 10 kpc in depth.}
\label{fig:fig2}
\end{figure*}%

The figure shows a few visually distinct regions. The particles launched into the jets constitute a thin, cylindrical region (often called the jet spine) of unshocked, cold gas (with the temperature the same as the ambient medium). This gas experiences some shocking all the way from the launching region to the jet head. The gas shocked in this way makes up the hot lobe that can be seen surrounding the jet itself. Finally, the action of the jet also launches a bow shock, which transitions from being strongly supersonic near the jet head to a sound wave in the perpendicular direction.

Visually inspecting the jet, we can surmise whether this jet is in the self-similar or ballistic phase. It is clear that the aspect ratio of the lobes (length vs. width) grows with time, whereas in the self-similar case it should remain constant. Furthermore, we can see that the jet increases in length by an approximately equal amount with each snapshot, indicating that the jet velocity is nearly constant (as it should be in the ballistic regime). More quantitatively, we find the power-law slope of the $L_\mathrm{j}-t$ dependence to be $0.9$, very near the ballistic value of $1$. 

The typical velocity of the jet head is found to be $\approx5500$ km/s. However, a ballistic jet should drill through the medium at exactly the launching velocity, which is $15000$ km/s in this case. Equally, one might wonder why is there significant shocking of the jet particles along the way from the launching region to the jet head, whereas we would expect all shocks to happen at the latter location (for ballistic jets). We have performed other simulations of ballistic jets, which we do not show here (since we focus on self-similar jets), where we find that this discrepancy is due to numerical resolution. In particular, we find that increasing the resolution leads to less shocking occurring inside the jet spine. The jet head velocity is consequently larger, and the jets thinner and denser. We find that simulated jets are close to being fully ballistic only at very high resolutions ($>10^5$ particles per jet).

\begin{figure*}
\includegraphics[width=1.01\textwidth, trim = 0 12 0 0]{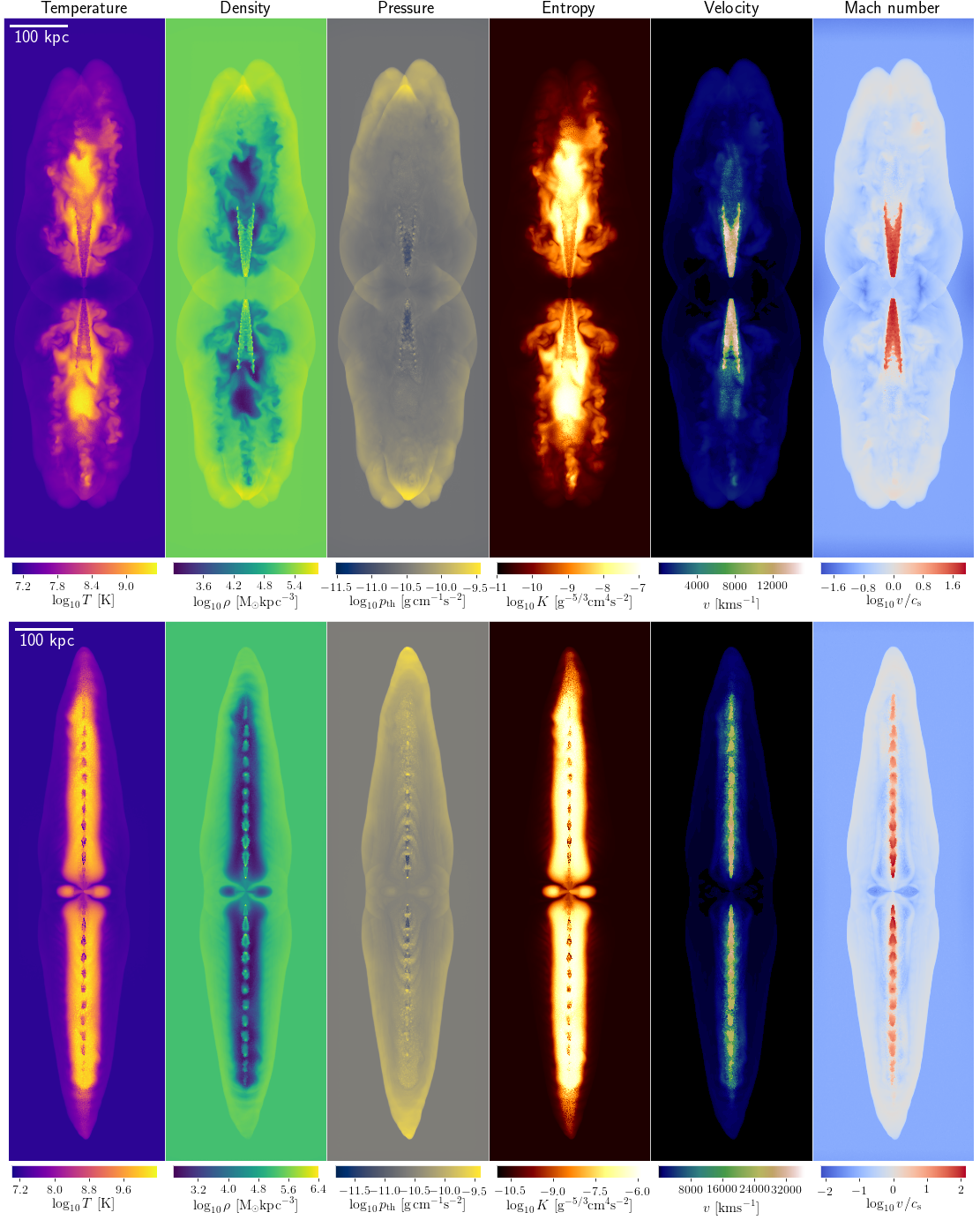}
\caption{Top: properties of our highest resolution jet simulation after $96$ Myr (with a mass resolution $m_\mathrm{gas}=5.73\times 10^3$ M$_\odot$, while other parameters correspond to our fiducial choice, given in the first row of Table \ref{tab:tab0}). Bottom: properties of a simulation with a slightly lower resolution, $m_\mathrm{gas}=1.81\times 10^4$ M$_\odot$, a smaller opening angle $\theta_\mathrm{j}=5\degree$ and a larger jet velocity $v_\mathrm{j}=30000$ km/s, after $48$ Myr. Each panel is 10 kpc in depth and shows different gas properties, as given by the titles and legends.}
\label{fig:fig3}
\end{figure*}%

\subsection{General properties of jets in the self-similar phase}

In the bottom panels of Fig. \ref{fig:fig1} we show a jet from a simulation with an equivalent set of parameters as the previously discussed ballistic jet simulation, but with a half-opening angle of $10\degree$ rather than $0\degree$. The jet power, $P_\mathrm{j}=10^{46}$ erg/s and the launching velocity $v_\mathrm{j}=15000$ km/s, yield the $L_1$ length-scale of $7.7$ kpc, on order of our launching region size. This means that jets should reach the self-similar phase almost as soon as they are launched. The $L_2$ length scale is $970$ kpc, so ambient pressure should not be important.

This jet is shorter and wider than the ballistic one, as one might intuitively expect. The central outflow of unshocked gas no longer forms a thin spine; this jet gas instead flows conically until it is recollimated at distances of $\approx50$ kpc (this is referred to as recollimation since all jets are initially collimated at small distances, e.g. \citealt{Park2019}, \citealt{Chatterjee2019}). The recollimation is driven by previously shocked jet gas, which constitutes a hot lobe surrounding the unshocked jet. The collimated gas then expands and collimates again in an oscillatory fashion; these spatially periodic recollimation shocks are expected theoretically (\citealt{Falle1991}, \citealt{Bamford2018}), have been found in other simulations (\citealt{vanPutten1996}, \citealt{Mizano2015}, \citealt{Hervet2017}, \citealt{Bodo2018}, \citealt{Gourgouliatos2018}, \citealt{Smith2019}) and could explain multiple hotspots observed in some radio jets (\citealt{Rees1978}, \citealt{Dreher1984}, \citealt{Hardcastle2003}).

In Fig. \ref{fig:fig2} we show a visualisation of our highest-resolution jet simulation, with the same set of parameters but simulated at a resolution of $m_\mathrm{gas}=5.73\times 10^3$ M$_\odot$ (316 times better than EAGLE resolution). The jet shown exhibits clear signs of a conical outflow of particles, which is collimated at the point where significant shocking begins to occur (visible by a change in temperature). Unlike the lower-resolution simulation, we see no sign of multiple recollimation shocks. Based on analytical expectations (Eqn. 26 from \citealt{Kaiser1997}), the initial recollimation shock should occur at $\approx 60$ kpc along the z axis for our jets. We find that the cone begins to lose coherence at roughly such a distance, albeit somewhat too far ($70-100$ kpc, difficult to pinpoint exactly since the outflow is shocked earlier in the centre of the cone than at the edges).

Due to instabilities, we find that the jet lobe does not exhibit a simple ellipsoidal shape. The lobe is also not very homogeneous in gas temperature (as well as other properties; see Fig. \ref{fig:fig3}), and this is likely the reason why there are no multiple recollimation shocks (recollimation requires uniform pressure from all sides). For the same reasons, the bow shock created by the jet exhibits some irregularities. The deviation from a smooth ellipsoidal shape (which we do find at lower resolutions, see Fig. \ref{fig:fig6}) could be explained in a few ways:
\begin{enumerate}
    \item Usage of a fairly low launching velocity means that the jet is heavy, and thus mixes with the ambient medium less easily (\citealt{Rossi2008}, \citealt{English2016}, \citealt{Donohoe2016}).
    \item Only the largest-wavelength Kelvin-Helmholtz instabilities are resolved in this simulation. In other words, it is possible that at even higher resolutions, a jet lobe would again be recovered, but a highly turbulent one. 
    \item The recollimation shock is expected to occur at a large distance from the jet origin (60 kpc). A lobe is not expected to exist until the jet has reached that distance. A lobe-like feature does appear and become more prominent from snapshot 2 to 3, and 3 to 4 in Fig. \ref{fig:fig2}. If the jet was allowed to become much larger than the distance to the recollimation shock, a clear lobe would likely be visible. This is supported by the equivalent lower-resolution jet also having a clear lobe only at later snapshots.
\end{enumerate} 

In the top panels of Fig. \ref{fig:fig3} we show several different gas properties in the high-resolution jet simulation after $96$ Myr of evolution. The conical outflow is constituted by cold, dense and fast-moving gas, with a high Mach number ($M\approx10-40$). This gas has a low entropy and is under-pressured compared to the jet lobe. This causes a collimation of the outflowing particles and subsequent shocking. 

The jet lobe, constituted from previously shocked jet particles, is of high temperature and low density (by a factor of up to $\approx300$ relative to the ambient medium, although this varies greatly within the lobes), as well as in pressure equilibrium. Its high entropy indicates that this gas has experienced significant shock heating. The typical velocity in the jet lobe is of order a few $\times$ $1000$ km/s, much lower than the launching velocity. Furthermore, although not visible on this plot, the velocity is not only in the z-direction. In particular, near the base of the jet, the lobe particles move in the direction opposite of the general jet direction, constituting the so-called backflow (\citealt{Lind1989}, \citealt{Rossi2008}, \citealt{Cielo2014}, \citealt{Mukherjee2018}).

The bow shock surrounding the jet lobes has a high density, indicating that it is constituted from particles swept up by the jet. It is mildly supersonic with Mach number of order $M\approx1.5$. Such Mach numbers have been found in deep X-ray observations of galaxy clusters, in the presence of weak shocks  (\citealt{Fabian2003}, \citealt{Forman2007}, \citealt{Snios2018}), and are thought to be one of the main ways that jets can heat the intra-cluster medium in an isotropic fashion (\citealt{Reynolds2001}, \citealt{Bruggen2007}, \citealt{Li2017}, \citealt{Weinberger2017}). Near the base of the jets, however, we find that the bow shock expands laterally at the sound speed. It is almost invisible on the entropy plot, indicating that it has experienced only mild shock heating, compared to both the lobe and the ambient medium near the jet head.

In the bottom panels of Fig. \ref{fig:fig3} we show a jet in the self-similar regime with a narrower opening angle ($\theta_\mr{j}=5\degree$) and higher launching velocity ($v_\mr{j}=30000$ km/s). The jet power is the same, while the mass resolution is relatively fine at $m_\mr{gas}=1.81\times10^4$ $\mr{M}_\odot$. This would generally yield a high-resolution jet, but usage of a large launching velocity (larger by factor of two compared to our standard choice) effectively reduces the jet resolution by a factor of four. This jet exhibits an interesting streak of more than a dozen recollimation shocks, owing to the narrow opening angle and large launching velocity. The Mach number in this case reaches values up to $M\approx100$. The large launching velocity results in very light and hot lobes with density/temperature contrasts of almost a factor of 1000, relative to the ambient medium.

\subsection{Evolution of the jet-inflated lobe length and width}

We now turn to comparing the length and radius of the lobes in our simulations to theoretical predictions of self-similar lobes discussed in Section \ref{sec:theory}. Here we show results for our standard-resolution simulation (i.e. the one shown in the bottom panels of Fig. \ref{fig:fig1}). We do this since for that simulation we have a relatively large number of snapshots (50), whereas we were only able to output four snapshots for the highest-resolution simulation (due to storage restrictions). We discuss the dependence of various properties on resolution in Section \ref{sec:mass_resolution}.

\begin{figure}
\includegraphics[width=0.99\columnwidth, trim = 0 15 0 0]{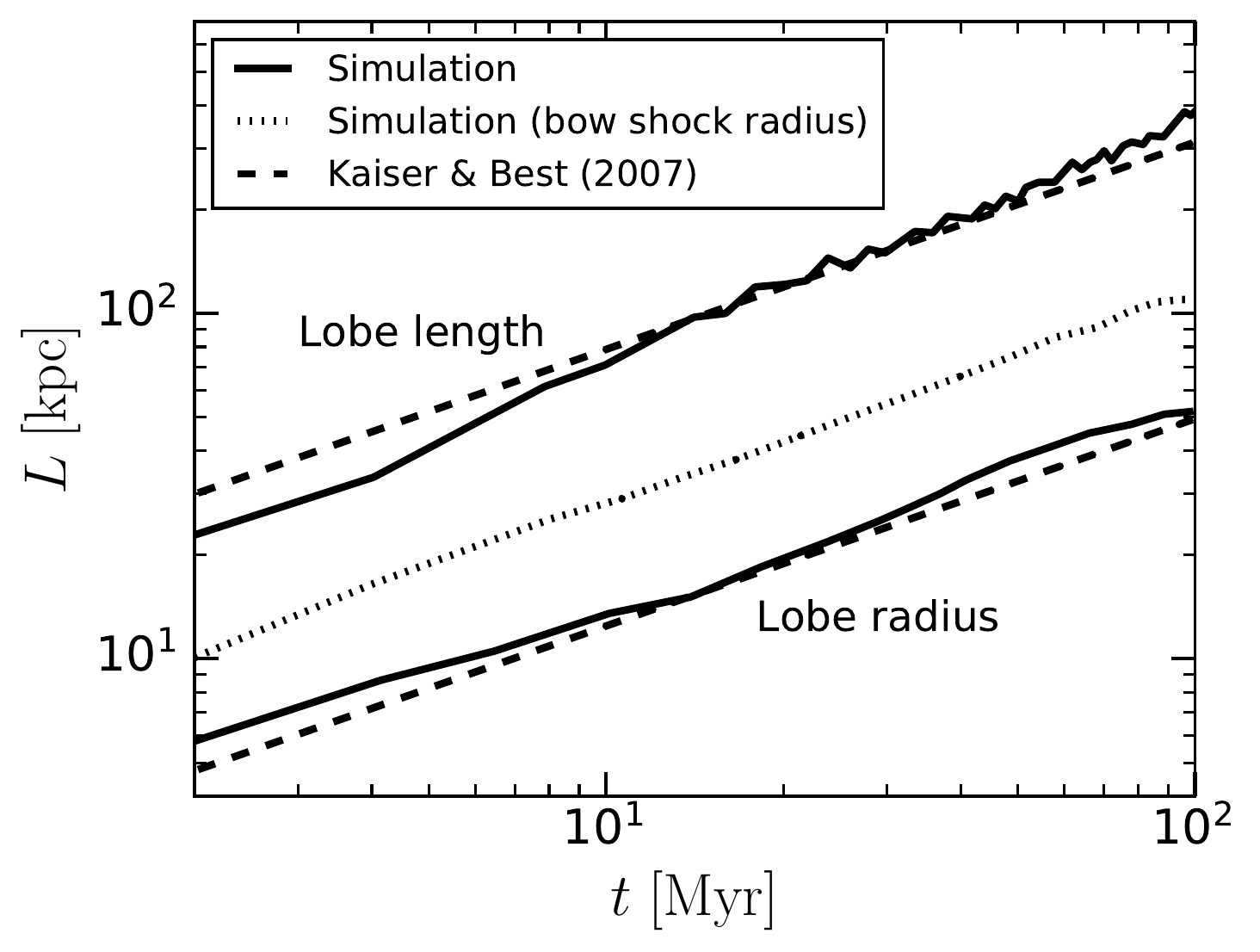}
\caption{Time evolution of the jet lobe length and radius, as well as the bow shock radius (solid and dotted lines, as marked), in our standard simulation (first row of Table \ref{tab:tab0}). These are compared to predictions from \protect\cite{Kaiser2007} for the lobe (dashed lines; Eqn. \ref{eq:self_similar_jet_length}).}
\label{fig:fig5}
\end{figure}

Fig. \ref{fig:fig5} shows the time dependence of jet/lobe length and lobe radius (at its widest point), as well as the bow shock radius, in our standard-resolution simulation. We also show the prediction based on approximate theoretical models of jets and lobes in the self-similar regime (Eqn. \ref{eq:self_similar_jet_length}, \citealt{Kaiser2007}). The agreement is clearly very good, especially at late times. This is true for both the lobe (and jet) length and the lobe radius, both of which have the same slope ($0.6$) in the time dependence. This means that the lobe has a constant aspect ratio, as predicted in the self-similar model, and its value is in agreement with the predictions. We find that the bow shock radius scales with time in the same way, which is also in agreement with the self-similar theory. We find that the lobe is slightly too short and too wide at $t<5$ Myr, although this is likely related to lower resolution at these times, since only a small fraction of particles have been launched into the jets.

\subsection{Jet, lobe and ambient medium energetics}

The question of how much energy is in which component, and in what form, is an interesting one in the context of jets as a feedback mechanism that plays a role in the formation and evolution of galaxies. We again focus on the energetics of our standard-resolution self-similar simulation, rather than the high-resolution one, due to more snapshots being available. We have performed similar analyses for the higher-resolution simulations and we find that the energetics is well converged, down to the level of a few per cent. The procedure for calculating the energies in the jets and lobes (here we group the jets into the category of 'lobe' for the purpose of simplicity), as well as the ambient medium, is described in Section \ref{sec:energetics_lengths}.

\begin{figure}
\includegraphics[width=0.99\columnwidth, trim = 0 15 0 0]{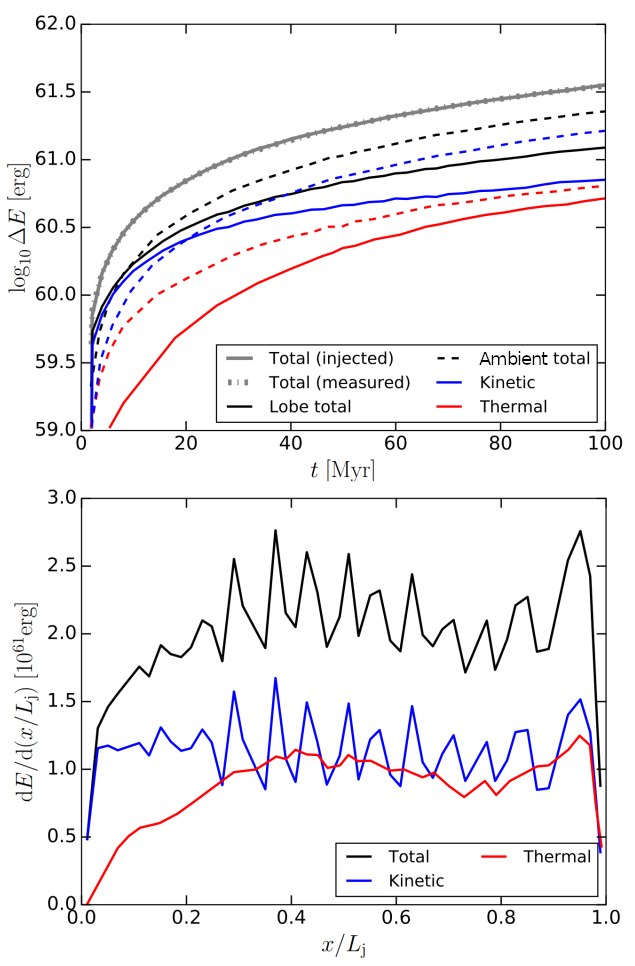}
\caption{Energetics of our standard jet simulation (see the first row of Table \ref{tab:tab0} for a list of parameters), at $t=100$ Myr. \textit{Top:} Evolution of the total energies in the jet lobe and ambient medium. Blue lines represent kinetic energy, while red lines are the thermal energy. The sums for the lobe and ambient medium are given by the black lines. The grey lines represent the total injected energy and the total change in the energy of the lobes and the ambient medium. These two lines overlap, showing that energy is numerically conserved. \textit{Bottom:} Kinetic, thermal and total energy densities per unit length along the jet axis, normalised by the total jet length.}
\label{fig:fig4}
\end{figure}

In the top panel of Fig. \ref{fig:fig4} we show the energetics of our standard-resolution simulation. The top panel shows the kinetic and thermal energies of the jet-inflated lobes and the ambient medium. The sum of all of these energies exactly matches the total injected energy, showing that energy is conserved. During the initial phase of jet launching, the ambient medium and the lobe are approximately equal in total energy. At later times, the lobe carries about a third of the energy, whereas the remainder has been transferred to the ambient medium. At late times this ratio is constant, as one would expect for a jet in the self-similar phase (since the volume ratios of the lobe and the region defined by the bow shock remain constant, due to the constant shape of both components).

For the jet lobes we find that they are dominated by kinetic energy initially, but by the time they are turned off (at the end of the simulation), an approximately equal amount of energy is in the thermal component. The ambient medium initially has roughly equal amounts of added kinetic and thermal energy, but at later times about two thirds of the added energy is in the kinetic component. More quantitatively, at $t=100$ Myr (the end of the simulation), the energy partition is as follows: 1) lobe kinetic - $20\%$, 2) lobe thermal - $14\%$, 3) ambient kinetic - $47\%$ and 4) ambient thermal - $19\%$. Across different simulations we find that these fractions can vary, but none of the components becomes negligible. 

In the bottom panel of Fig. \ref{fig:fig4} we show the energy density per unit length in the lobe, including the jet itself (normalised by the length of the jet), at the end of the same simulation. Note that this is the energy density per unit length in slices perpendicular to the jet axis, and not just the energy density per unit length exactly along that axis. The kinetic energy density is roughly constant along the entire length, but shows signs of oscillations around the constant component. These oscillations are a result of the multiple recollimation shocks. Thermal energy is initially negligible, but reaches about the same density as the kinetic component at one third the jet length. The total energy density first rises, reaches a slight peak around half the jet length, and then falls - this is a result of the ellipsoidal shape. However, at the very end of the jet, both the thermal and kinetic energies reach a peak. This is likely a feature of the terminal shock.

\subsection{Comparison with previous simulations}
\label{sec:comparison_sims}

Here we will compare our results with previous hydrodynamical simulations of AGN jets, all of which were performed using grid-based codes. While there is an extensive literature of such simulations, most of these include and study the effects of more exotic physics that we do not include, such as magnetic fields (e.g. \citealt{Hardcastle2014}, \citealt{Tchekhovskoy2016}, \citealt{Mukharjee2020}), radiative cooling (e.g. \citealt{Blondin1990}, \citealt{Stone1997}, \citealt{Guo2018}) and cosmic rays (e.g. \citealt{Guo2011}, \citealt{Ehlert2018}, \citealt{Yang2019}). We thus do not compare with such studies. We also restrict our comparison to studies that launch jets into a constant-density ambient medium, since this is the focus of our study. We do not, however, restrict ourselves to comparisons with non-relativistic jet studies. We do this since the majority of the literature has included relativistic effects, and since differences between classical and relativistic jets (or more accurately, the lobes they inflate) should be minor (\citealt{Kaiser2007}).

The above restrictions leave only a few studies that are comparable with out study. We begin by comparing with the results of \cite{Falle1991}, who also presented the first analytical model of self-similar jet lobe expansion. They performed 3D simulations of classical jets launched into an ambient medium of fixed density. They did not use a finite (non-zero) opening angle (unlike our study), since they argue that this requires the jet launching region to be well resolved. Instead they used a zero opening angle and a relatively small internal Mach number, which should be similar to using a non-zero opening angle in combination with a large internal Mach number ($M\gtrsim10)$. We independently confirm this to be the case, although we do not show the results of those simulations here.

These two different set-ups lead to a similar outcome for the following reasons. In the finite-opening angle case, the outflowing jet gas has a conical geometry as a direct consequence of how it was set up. The gas is eventually recollimated once the pressure in the jet cone becomes lower than the pressure of the lobe gas. In the case of a zero-opening angle and small internal Mach number, the outflowing jet gas is instead hot enough that it expands on account of thermal pressure. It recollimates for the same reason as the finite-opening-angle case. We note, however, that the relation between an 'effective' opening angle (or aspect ratio of the lobes) and the finite Mach number is not obvious. The value \cite{Falle1991} used is $M=5$, which they find corresponds to $\theta_\mr{j}=13\degree$. They found that their jets inflate lobes that expand self-similarly, i.e. with a constant aspect ratio and whose length evolves with time as $L\propto t^{3/5}$. Both of these findings are in good agreement with our results (\ref{fig:fig5}). For comparison, our standard simulations have a finite opening angle of $10\degree$ and an internal Mach number of $M\approx20$.

\cite{Komissarov1998} performed simulations that are more directly comparable to ours, since they used finite opening angles ($\theta_\mr{j}=5-20\degree$) and very large internal Mach numbers. These simulations were also 3D, and included both the classical and relativistic variety of jets. They found that the latter are very similar to the former in qualitative behaviour, so we do not differentiate between the two for the purpose of this comparison. With their set-up, they also found jets whose lobes reach the self-similar regime at late times, with $L\propto t^{3/5}$. However, they found that this regime is not reached as soon as the jet is much larger in length (e.g. several times) than the transition length-scale $L_1$ that separates the ballistic and self-similar evolutionary period (Eqn. \ref{eq:first_scale}). Instead, the transition is complete once the \textit{width} of the jet-inflated lobe is a few times larger than $L_1$. They also found that the transition is very prolonged, and there is no obvious break in the velocity, aspect ratio or jet length (i.e. the jet length does not suddenly change from scaling linearly with time to a $\propto t^{3/5}$ scaling around $L_1$). Our findings are also in good agreement with these results. From Fig. \ref{fig:fig5} we see that there is no sudden transition at around $L_1$, which is $\approx8$ kpc in our case. Instead, the jet is too short but also too wide (and therefore too stumpy) compared to the self-similar prediction. By the time the width of the lobe is $\approx15$ kpc $=(2-3)\times L_1$, the jet has, however, reached the self-similar regime. Both the length and the radius of the lobe begin to agree with the self-similar prediction at the same time. This is in good agreement with the finding of \cite{Komissarov1998}, who argue that this transition is complete once the aspect ratio has increased enough to reach the value it should have in the self-similar regime.

Finally, we briefly compare with some of the newer work on hydrodynamical AGN jets by \cite{Krause2012}. They used finite opening angles in a 2.5D simulation with a constant-density ambient medium. Their jets are very similar in appearance to those of \cite{Komissarov1998}, both of which show more mixing than our jets. It is hard to compare our results quantitatively (since \citealt{Krause2012} do not show results on the evolution of the lobe length and radius), but we note that \cite{Krause2012} also find a distinct transition that occurs once the jets reach the $L_1$ length-scale.



\section{Results: comparison of jets with varying parameters}
\label{sec:parameter_study}

In this section we compare the properties of jets simulated with different parameters. As in the previous section, all of these results concern our constant-density ambient medium case, with the exception of the results discussed in Section \ref{sec:power_law}.

The parameters we vary here are jet power, half-opening angle, launching velocity and mass resolution. The standard choice and ranges of variations are given in Table \ref{tab:tab0}, and we also repeat the former here: $m_\mathrm{gas}=1.81\times 10^5$ M$_\odot$, $P_\mathrm{j}=10^{46}$ erg/s, $\theta_\mathrm{j}=10\degree$ and $v_\mathrm{j}=15000$ km/s. When varying any one of the parameters, we keep all other parameters fixed. For each variation, we have performed 5 simulations with different parameter choices (with the exception of mass resolution, where we have performed 8 simulations in total, in order to probe the full range form very low to very high-resolution jets).

\begin{figure*}
\includegraphics[width=0.97\textwidth, trim = 0 12 0 0]{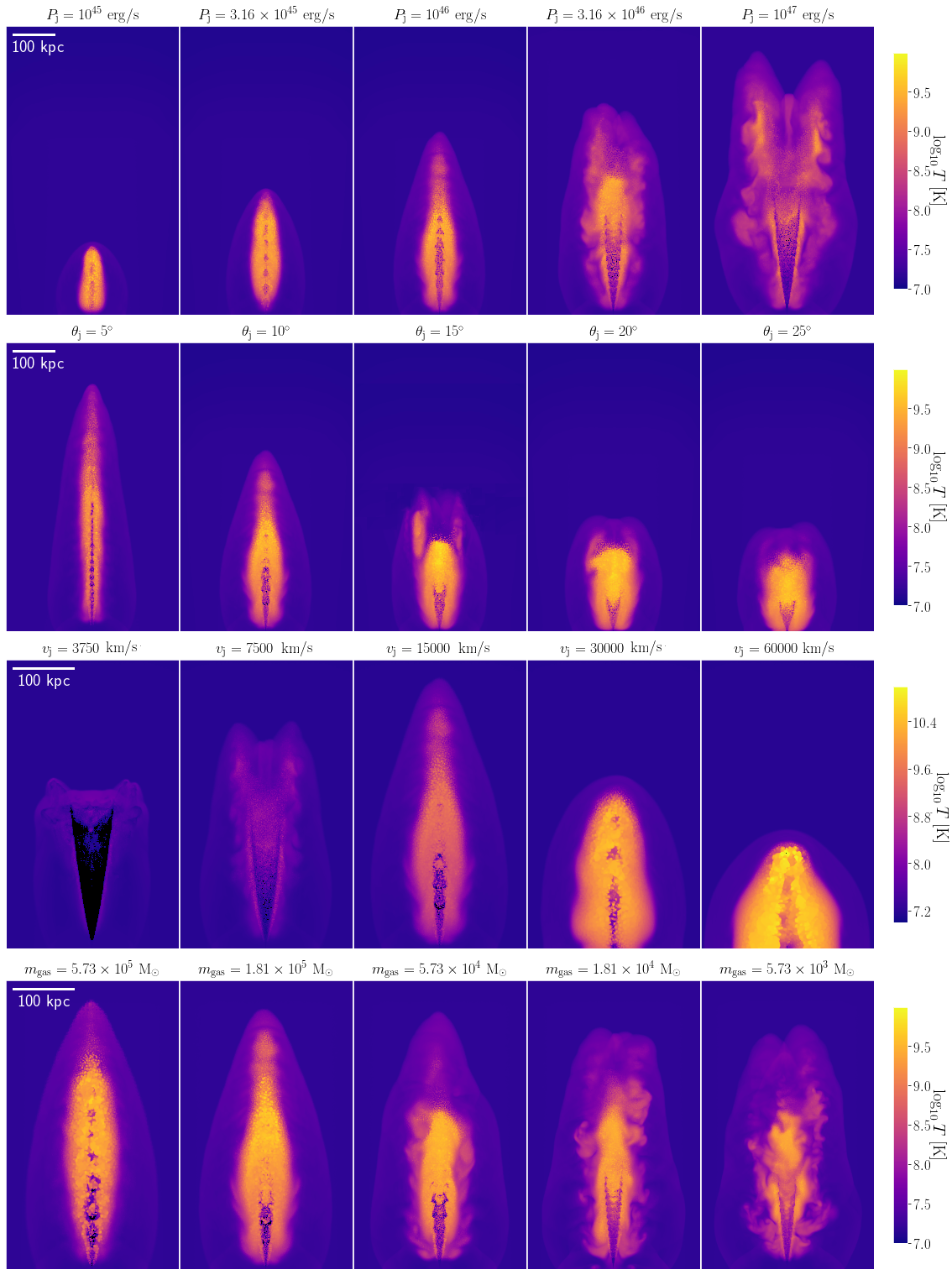}
\caption{Comparison of jets simulated with different parameters (see Table \ref{tab:tab0} for the range of variations), after $100$ Myr of evolution. The standard set of parameters is: $P_\mathrm{j}=10^{46}$ erg/s, $\theta_\mathrm{j}=10\degree$, $v_\mathrm{j}=15000$ km/s and $m_\mathrm{gas}=1.81\times 10^5$ M$_\odot$. When varying one parameter, all others remain fixed. From top to bottom, each row contains variations of: jet power, half-opening angle, launching velocity and mass resolution. The colours show gas temperature, according to the colour bars. The panels are 10 kpc in depth.}
\label{fig:fig6}
\end{figure*}

We varied the jet power by factors of $\sqrt{10}\approx3.16$ between $P_\mr{j}=10^{45}$ erg/s and $P_\mr{j}=10^{47}$ erg/s, the half-opening angle in increments of $5\degree$ from $\theta_\mr{j}=5\degree$ to $\theta_\mr{j}=25\degree$, the jet launching velocity by factors of 2 from $v_\mr{j}=3750$ km/s to $v_\mr{j}=60000$ km/s, and finally the mass resolution by factors of $\sqrt{10}\approx3.16$ from $m_\mr{gas}=1.81\times10^7$ $\mr{M}_\odot$ to $m_\mr{gas}=5.73\times10^3$ $\mr{M}_\odot$. Fig. \ref{fig:fig6} shows the visualisation of jets from all of these simulations after $100$ Myr of evolution. In Fig. \ref{fig:fig7} we show the time evolution of jet/lobe lengths for all four cases of parameter variations.

\subsection{Varying the jet power}

The main effect of increasing the jet power, as is visible in the first row of Fig. \ref{fig:fig6}, is the lengthening of the jets. The typical temperature of the jet-inflated lobes remains the same. This is because the launching velocity is kept constant ($v_\mathrm{j}=15000$ km/s). At lower jet powers, we find a roughly ellipsoidal shape for jet lobes, and multiple recollimation shocks. At higher jet powers, instabilities destroy this ellipsoidal lobe shape. This is possibly a result of these higher-power jets being better resolved due to more particles being injected into the jets. The lobe disruption results in a much more complex structure, similar to our high-resolution simulation with standard jet power (Fig. \ref{fig:fig2}). In our highest jet power simulation, the lobe (and therefore the bow shock) takes on a horned shape at the jet head, which is also similar to the high-resolution simulation shown in Fig. \ref{fig:fig2}. This is a result of the transition length scale $L_1$ (that determines when the jet goes from the ballistic to the self-similar regime) being larger at higher jet powers (Eqn. \ref{eq:first_scale}). As a result, this jet is ballistic for a longer portion of its lifetime shown here.

From the top left panel of Fig. \ref{fig:fig7}, we can see that the jet/lobe length evolution is self-similar for the three lower jet powers, and in fairly good agreement with theoretical predictions. At high jet powers, however, the slope of the time dependence is closer to $0.7$ than the self-similar value of $0.6$. This is possibly due to the already-mentioned instabilities causing some deviation from the self-similar model. Alternatively, the time dependence in these two highest power simulations could be interpreted as showing the gradual transition from a ballistic phase with $L_\mr{j}\propto t$ at early times to a self-similar phase with $L_\mr{j}\propto t^{0.6}$ at late times.

\subsection{Varying the half-opening angle}

From the second row of Fig. \ref{fig:fig6}, we see that jets become shorter as their half-opening angle increases, also accompanied by an increase in the aspect ratio of the lobes. This is followed by the disappearance of multiple recollimation shocks. Instead, the outflow takes the shape of a simple cone with a single recollimation shock. The ballistic phase of the jet can be seen in the shocked jet gas near the bow shock. In the two lower-angle simulations, this gas takes the form of a thin strip, whereas at larger opening angles we see evidence of a horned feature. From the top right panel of Fig. \ref{fig:fig7} we can see that all jets, except that for $\theta_\mathrm{j}=5\degree$, agree well with the self-similar prediction of jet and lobe length evolution. This is because the $\theta_\mathrm{j}=5\degree$ jet is ballistic for a fairly long time.

\subsection{Varying the launching velocity}

\begin{figure*}
\includegraphics[width=0.99\textwidth, trim = 0 15 0 0]{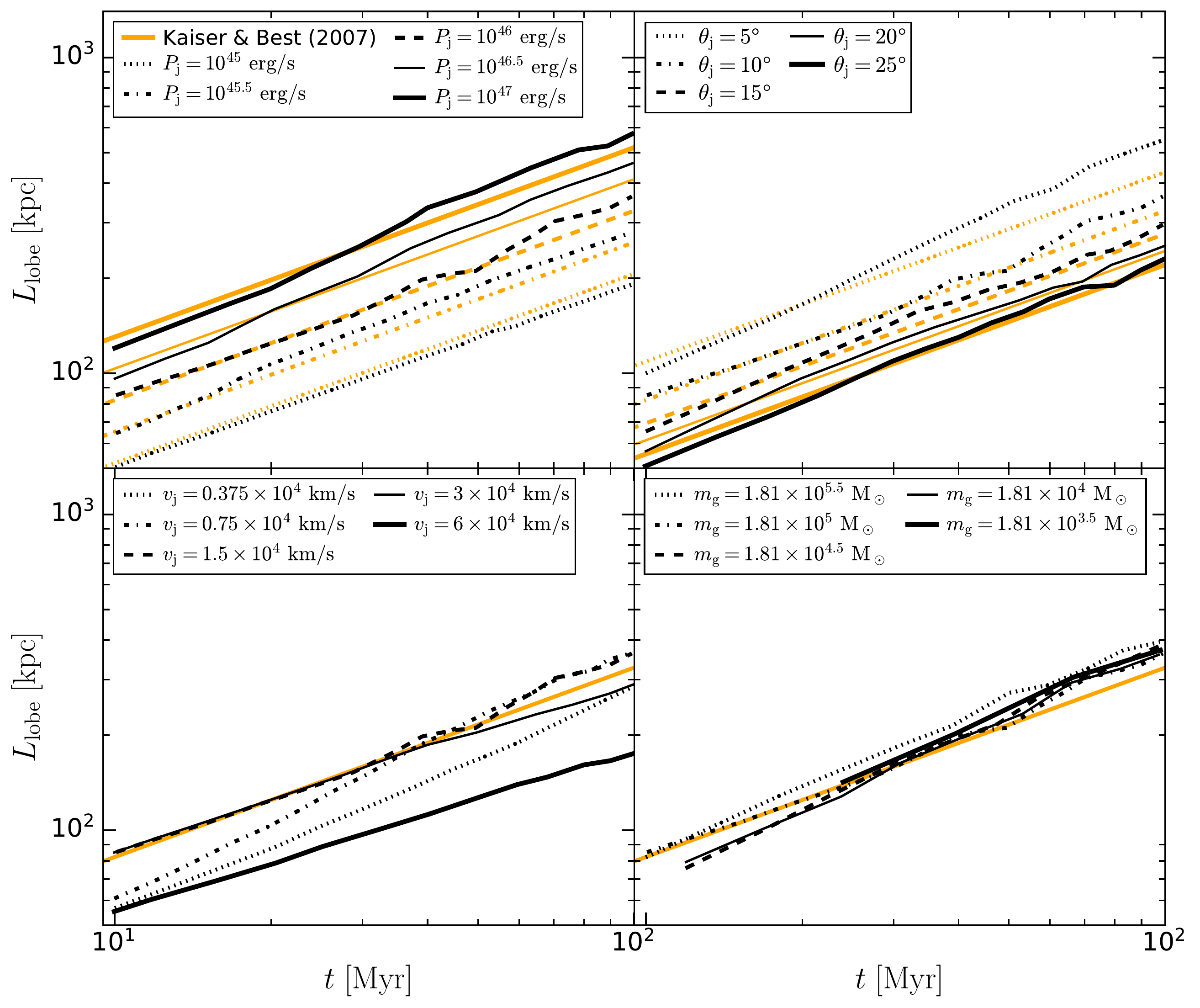}
\caption{Jet and lobe length as a function of time for simulations with various parameters. The standard set of parameters is: $P_\mathrm{j}=10^{46}$ erg/s, $\theta_\mathrm{j}=10\degree$, $v_\mathrm{j}=15000$ km/s $m_\mathrm{gas}=1.81\times 10^5$ M$_\odot$, and the range of variations is given in Table \ref{tab:tab0}. Each panel represents a different variation of parameters, as visible in the legends (top left - jet power, top right - half-opening angle, bottom left - launching velocity, bottom right - mass resolution). Orange lines represent theoretical predictions from \protect\cite{Kaiser2007}. Where we show only one line, there should be no change of jet length with the parameter being varied.}
\label{fig:fig7}
\end{figure*}%

The third row of Fig. \ref{fig:fig6} shows that the jet-inflated lobes change shape significantly, as well as becoming hotter, as we increase their launching velocity. The latter finding is expected, since there is more kinetic energy per particle available to be thermalised (due to constant total energy and a smaller number of particles). According to the self-similar model, changing the launching velocity of a jet should not result in any change in the jet/lobe length (see Eqn. \ref{eq:self_similar_jet_length}). However, the launching velocity can affect when the evolution changes from ballistic to self-similar. We find that jets become shorter, relative to the $v_\mathrm{j}=15000$ km/s case, as we either increase or decrease the launching velocity. 

The shortening at higher $v_\mathrm{j}$ is unexpected since the ballistic phase should in this case be even shorter. This effect is possibly due to lower resolution from having a smaller number of particles in the jet (preventing a coherent central jet spine from traveling to larger distances). However, based on our results with varying mass resolution, down to very low resolutions of $\approx500$ particles per jet (see next subsection), these high-velocity jets should still reach the same size as our standard velocity case. According to the simulations by \cite{English2016} and \cite{Li2018}, jet-inflated lobes in simulations with higher launching velocities should be lighter and wider at the base, due to a stronger backflow of gas near the terminal shock. We find that the two lobes are indeed wider at the base, to the point of the top and bottom lobes merging. This is possibly either due to a poorly-resolved backflow, or thermal expansion. Since the self-similar model of jet evolution takes the latter into account, but not the former, we conclude that these stubbier jets at very high velocities are likely due to a backflow, in agreement with previous simulations. We have performed other simulations with similarly large velocities, including 2D simulations with very high resolutions and near-relativistic velocities. We find the presence of strong backflows in all cases, causing the merging of the two lobes and a relatively stunted lobe.

We now move on to the lower-velocity cases. In the $v_\mathrm{j}=7500$ km/s simulation, the jet is slightly shorter than the $v_\mathrm{j}=15000$ km/s case, again due to instabilities being resolved better, which is itself a consequence of a larger number of particles being launched into the jets with lower velocities. However, at these velocities we are also approaching the ballistic limit, where the jet head velocity cannot be larger than the launching velocity. The ballistic nature is visible in the central jet outflow, which shows no sign of multiple recollimation shocks, but instead takes on a conical shape. Between this cone and the bow shock is a significant amount of shocked gas, indicating that this jet is not fully ballistic. The shocked gas, and therefore the bow shock, takes on a horned morphology. This can be traced to the fact that gas in the centre of the conical outflow is shocked at earlier times/smaller distances than that at the edges of the cone, resulting in a horned feature (if viewed as a slice; in 3D this feature is a ring). Similar features have been found in other simulations (\citealt{Omma2004}, \citealt{Matsumoto2019}, \citealt{Talbot2020}). In the $v_\mathrm{j}=3750$ km/s case, the jet is fully ballistic, which is visible by the conical outflow extending all the way to the bow shock. This jet is shorter because its launching velocity is smaller than the self-similar jet head velocity we expect in this case ($\approx5000$ km/s). While this is a small difference, there is invariably some shocking affecting the ballistic jet, causing its effective velocity to be less than $3750$ km/s.

From the bottom left panel of Fig. \ref{fig:fig7} we see the evolution of jet lengths with time for all five cases. The  $v_\mathrm{j}=7500$ km/s jet length agrees very well with our fiducial choice $v_\mathrm{j}=15000$ km/s, but only at late times (once it has reached the self-similar phase). The higher velocity jet with  $v_\mathrm{j}=30000$ km/s is similar to our fiducial case at early times, but slightly deviates from the theoretical prediction later on, possibly when a backflow begins to operate. Our highest velocity jet is in the self-similar regime at all times shown here, but with a lower normalisation, probably due to a backflow.

\subsection{Varying the mass resolution}
\label{sec:mass_resolution}

As is visible in the last row of Fig. \ref{fig:fig6}, the mass resolution does not affect the lobe lengths and widths significantly, which is encouraging. If anything, the jet lengths decrease slightly with resolution, showing that high resolutions are not necessary in order for jets to be able to deposit their energy at large distances$-$a conclusion relevant for cosmological simulations. The jets are somewhat shorter at high resolutions since the Kelvin-Helmholtz instability is better resolved, and this instability increases the effective inertia of the lobes by mixing them with the ambient medium. We find that the lobes take the expected ellipsoidal shape at lower resolutions, whereas such a shape is only beginning to form at high resolutions. At the same time, due to the lack of a coherent lobe, there is only a single recollimation shock at high resolution, at the end of the conical outflow, whereas we see multiple such shocks at low resolutions. We posit that this is due to the lobe no longer being as uniform in pressure (due to instabilities), which prevents the outflowing jet gas from being uniformly collimated at regular intervals. We find that the jets and lobes have not fully converged in structure by our highest resolution level. At even higher resolutions, smaller-scale vortices caused by the Kelvin-Helmholtz instability would possibly help to recover a more ellipsoidal lobe.

At all resolutions, we can see that the main ellipsoidal lobe does not extend all the way to the bow shock, but rather out to roughly three quarters of the distance to it. The lobe is instead connected to the bow shock by a thin strip of shocked jet particles. These are most easily visible in the three lower resolution snapshots, and especially at $m_\mr{gas}=1.81\times10^5$ $\mr{M}_\odot$. These particles are among the first that were launched into the jet, and they are the relic from its ballistic phase, with the self-similar lobe being built up in its wake. At higher resolutions, these particles are not easily visible because they were likely mixed with the ambient medium. From the bottom right panel of Fig. \ref{fig:fig7}, we can see that despite the qualitative differences in jet morphology, the evolution of jet lengths is in good agreement with the self-similar theory at all resolution levels. Our simulated jets show a slightly steeper evolution with time than expected from self-similar models, more consistent with $L_\mr{j}\propto t^{0.7}$ than $L_\mr{j}\propto t^{0.6}$.

\subsection{Varying the mass resolution - implications for cosmological simulations}

\begin{figure*}
\includegraphics[width=0.99\textwidth, trim = 0 5 0 0]{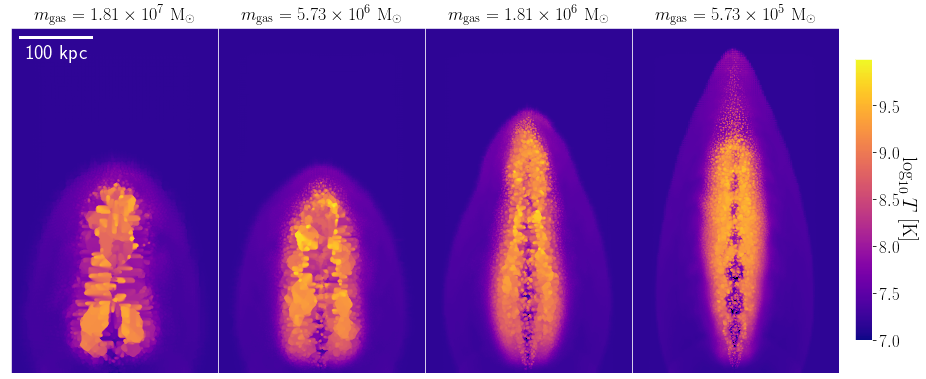}
\caption{Comparison of jets simulated with varying numerical resolutions (see top of each panel, these are typical resolutions employed in cosmological simulations of galaxy clusters or large-volume cosmological simulations). The number of particles in each jet varies from $\approx450$ to $\approx14000$. Other jet parameters used are the same as the fiducal case, see Table \ref{tab:tab0}. The panels show the jets after $100$ Myr of evolution. The colours show gas temperature, according to the colour bar. The panels are 10 kpc in depth.}
\label{fig:figA}
\end{figure*}%

In the bottom row of of Fig. \ref{fig:fig6} we showed how our jets vary visually across 5 resolution levels, separated by factors of 3.16. However, the lowest-resolution case shown ($m_\mr{gas}=5.73\times10^5$ $\mr{M}_\odot$) is itself better than the EAGLE cosmological simulation (\citealt{Schaye2015}) by a factor of 3.16, and it has a relatively high number of particles per jet, $\approx14000$. As a result, the comparisons shown in those figures, while interesting, are not necessarily relevant for cosmological simulations.

In order to probe more poorly-resolved jets, such as those likely to occur in large-scale cosmological simulations, we extend our analysis down to even lower resolutions. In Fig. \ref{fig:figA} we show the equivalent visualisations of jets as in the bottom row of Fig. \ref{fig:fig6}, but for resolutions down to ten times worse than EAGLE, i.e. $m_\mr{gas}=1.81\times10^7$ $\mr{M}_\odot$. The highest-resolution simulation shown is the one with $m_\mr{gas}=5.73\times10^5$ $\mr{M}_\odot$. As we can see, reducing numerical resolution results in a shortening of the jets and lobes they inflate. This is likely due to spherical averaging from SPH effects, as the total number of the jet particles approaches the number of particles expected in a single SPH smoothing kernel ($\approx50$). Despite the shortening, the jet (lobe) length is still within $15$\% of the self-similar prediction. This is true even for the lowest-resolution case shown here, which has only $\approx450$ particles launched per jet. We have simulated cases down to $100$ particles - these jets appear very spherical due to SPH averaging, with their jet and lobe components blended together. They are also significantly (factor of $\approx2$) shorter than the self-similar prediction.

Our worst-resolution simulation, with $m_\mr{gas}=1.81\times10^7$ $\mr{M}_\odot$, matches the resolution of some simulations of galaxy clusters (e.g. \citealt{Dubois2010}, \citealt{Martizzi2014}, \citealt{Richardson2016}, \citealt{Hahn2017}) and of cosmological simulations that are large enough in volume to contain many galaxy clusters (e.g. \citealt{Bocquet2016}, \citealt{Kaviraj2017}, \citealt{Pillepich}). This means that jet episodes with our fiducial power ($P_\mr{j}=10^{46}$ erg/s) and duration ($T_\mr{j}=100$ Myr) are likely to be well-converged in such simulations, at least in terms of basic properties. 

State-of-the-art zoom-in simulations of galaxy clusters often have resolutions significantly better than $m_\mr{gas}\approx10^7$ $\mr{M}_\odot$ (e.g. $m_\mr{gas}\approx10^6$ $\mr{M}_\odot$ in \citealt{Barnes2017}, \citealt{Bahe}, or $m_\mr{gas}\approx10^5$ $\mr{M}_\odot$ for a $M_\mr{h}=10^{14}$ $\mr{M}_\odot$ halo in \citealt{Tremmel2019}), so even weaker jet episodes are likely to be resolved at a basic level in these simulations ($P_\mr{j}=10^{44}$ erg/s in a galaxy cluster with $M_\mr{h}=10^{15}$ $\mr{M}_\odot$ or $P_\mr{j}=10^{43}$ erg/s in one with $M_\mr{h}=10^{14}$ $\mr{M}_\odot$). Modern cosmological simulations with box sizes of $L\approx100$ Mpc have a resolution of $m_\mr{gas}\approx10^6$ $\mr{M}_\odot$ (\citealt{Vogelsberger2014}, \citealt{Schaye2015}, \citealt{Pillepich}), and thus our numerical scheme can correctly model jets in such simulations at least down to systems with $M_\mr{h}=10^{14}$ $\mr{M}_\odot$, and possibly $M_\mr{h}=10^{13}$ $\mr{M}_\odot$.

Small zoom-in cosmological simulations often aim to reproduce individual Milky-Way type systems (\citealt{Kim2014}, \citealt{Hopkins2014}, \citealt{Sawala2015}, \citealt{Fattahi2016},\citealt{Grand2016}), or host multiple such systems in group environments within a $L=25$ Mpc box (\citealt{Crain2015}, \citealt{Tremmel2017}, \citealt{Dubois2021}). There is some evidence that jet feedback could be important even in such lower-mass systems (\citealt{Kaviraj2015}, \citealt{Singh2015}, \citealt{Olguin2020}). These simulations do not host any galaxy clusters, and can thus be simulated at much higher resolutions of $m_\mr{gas}\approx10^3-10^5$ $\mr{M}_\odot$, depending on the type of simulation. While this is very low, the halo masses in these systems are lower by similar factors (relative to a resolution of $m_\mr{gas}\approx10^6-10^7$ $\mr{M}_\odot$). The black holes hosted by these systems are less massive by even larger factors, due to the approximate $M_\mr{BH}\propto M_\mr{h}^{1.5}$ scaling (\citealt{Croton2006}, \citealt{Bandara2009}). In combination with shorter jet episodes of $1-10$ Myr (\citealt{Guo2012fermi}, \citealt{Garofalo2018}, \citealt{Davis2022}), instead of few 10s or up to 100 Myr (\citealt{Konar2006}, \citealt{Machalski2007}, \citealt{Mahatma2020}), it is possible that jet episodes in these systems could be less energetic by factors of up to $10^6$. It is unclear if simulations with $m_\mr{gas}\approx10^3-10^5$ $\mr{M}_\odot$ can correctly include such jet episodes, in the context of our numerical scheme.

As an example, we take a jet episode with $P_\mr{j}=10^{42}$ erg/s, lasting $5$ Myr. This jet power and duration correspond to the medians for jets in radio AGN (\citealt{Mezcua2019}, \citealt{Davis2022}), and the total energy matches the episode that likely recently occurred in the Milky Way (\citealt{Guo2012fermi}, \citealt{Predehl2020}). Such an episode is $2\times10^5$ times less energetic than the standard one we simulate. Assuming the same launching velocity of $v_\mr{j}=15000$ km/s, the mass resolution required to resolve each of these jets with $450$ particles (the resolution we have shown to be sufficient) would be $m_\mr{gas}\approx10^2$ $\mr{M}_\odot$. This is about an order of magnitude smaller than achievable with the highest-resolution cosmological simulations performed thus far (\citealt{Hopkins2018}). A convenient workaround, but still a physically relevant one, is to launch jets in lower-mass systems with lower velocities. As an example, $v_\mr{j}=1500$ km/s in a Milky-Way type system would produce a jet that is similarly underdense and hot compared to its ambient medium as in our simulations (factor of $\approx100$), due to a lower sound speed in the gas halo of the Milky Way ($c_\mr{s}\approx150$ km/s with $T_\mr{vir}=10^6$ K). This would imply that our example jet episode with $P_\mr{j}=10^{42}$ erg/s and $T_\mathrm{j}=5$ Myr could realistically be simulated with $m_\mr{gas}\approx10^4$ $\mr{M}_\odot$ or lower. This is in reach for modern cosmological simulations.

Overall, for our standard case, we find jets of very similar lengths and shapes (and therefore energetics, as the two are closely connected in self-similar models of jet lobes) across a very large range of numerical resolution: from $m_\mr{gas}=1.81\times10^7$ $\mr{M}_\odot$ down to $m_\mr{gas}=5.73\times10^3$ $\mr{M}_\odot$, a factor of 3160 difference. The implications of these findings, as discussed above, is that jets are likely to be correctly simulated at a basic level in SPH codes in all systems where they may be relevant. A caveat to this is that the highest resolutions achievable may need to be employed. Otherwise, low-power jet episodes will be poorly resolved. However, this is not necessarily a large issue, since it is likely that stronger episodes are more relevant in terms of feedback on galaxy formation. 

An additional caveat is that one may need to carefully vary the jet launching velocity from system to system to ensure that the jets are sufficiently resolved, but also that sufficiently strong shocks occur in these systems (as well as sufficiently large density and temperature contrasts). For this reason, constant-velocity schemes with jet launching velocities of $\approx10^4$ km/s (e.g. \citealt{Dubois2012}) will likely lead to very poorly resolved jets in any system with a dark matter halo mass below $M_\mr{h}\approx10^{14}$ $\mr{M}_\odot$, at resolutions comparable to EAGLE ($m_\mathrm{gas}\approx10^6$ $\mathrm{M}_\odot$). We instead suggest that scalings such as $v_\mr{j}= A M_\mr{h}^\gamma$ or $v_\mr{j}= A M_\mr{BH}^\gamma$ may be more appropriate. The normalization and slope can be chosen in such a way that the ratio between typical temperature of shocked jet gas (i.e. the lobe temperature) and that of the ambient medium is roughly constant regardless of the mass of the system (and of order e.g. 100, corresponding to jets that are supersonic by a factor of 10).

\subsection{Power-law ambient medium density profiles}
\label{sec:power_law}

Here we will discuss some results of simulations where jets were launched into gaseous atmospheres with power-law density profiles ($\rho\propto r^{-\alpha}$), rather than into a constant-density ambient medium. The set-up for these simulations is described towards the end of Section \ref{sec:setup}. We test cases with $\alpha=0.5$, $\alpha=1$ and $\alpha=1.5$, all of which should, in principle, feature the self-similar phase of jet evolution (if various parameters are chosen correctly). Given the $\beta$ profile we have chosen (Eqn. \ref{eq:beta_profile}), outside the small core, the density profile is $\rho(r)=\rho_{0,\beta}r^{-\alpha}$. The normalization $\rho_{0,\beta}$ depends on $\alpha$ in our set-up, since we have chosen to keep the mass of the gaseous halo constant within the virial radius of the external NFW potential. This means that an increase in $\alpha$ is also followed by an increase in $\rho_{0,\beta}$. An alternative would have been to keep the normalization constant, but to instead track the evolution of the jets over larger distances. We chose against this, and instead chose to keep the jets confined within the extent of gaseous haloes of the same size. We did this in order to be able to compare the jets at a similar size scale.

\begin{figure*}
\includegraphics[width=0.99\textwidth, trim = 0 5 0 0]{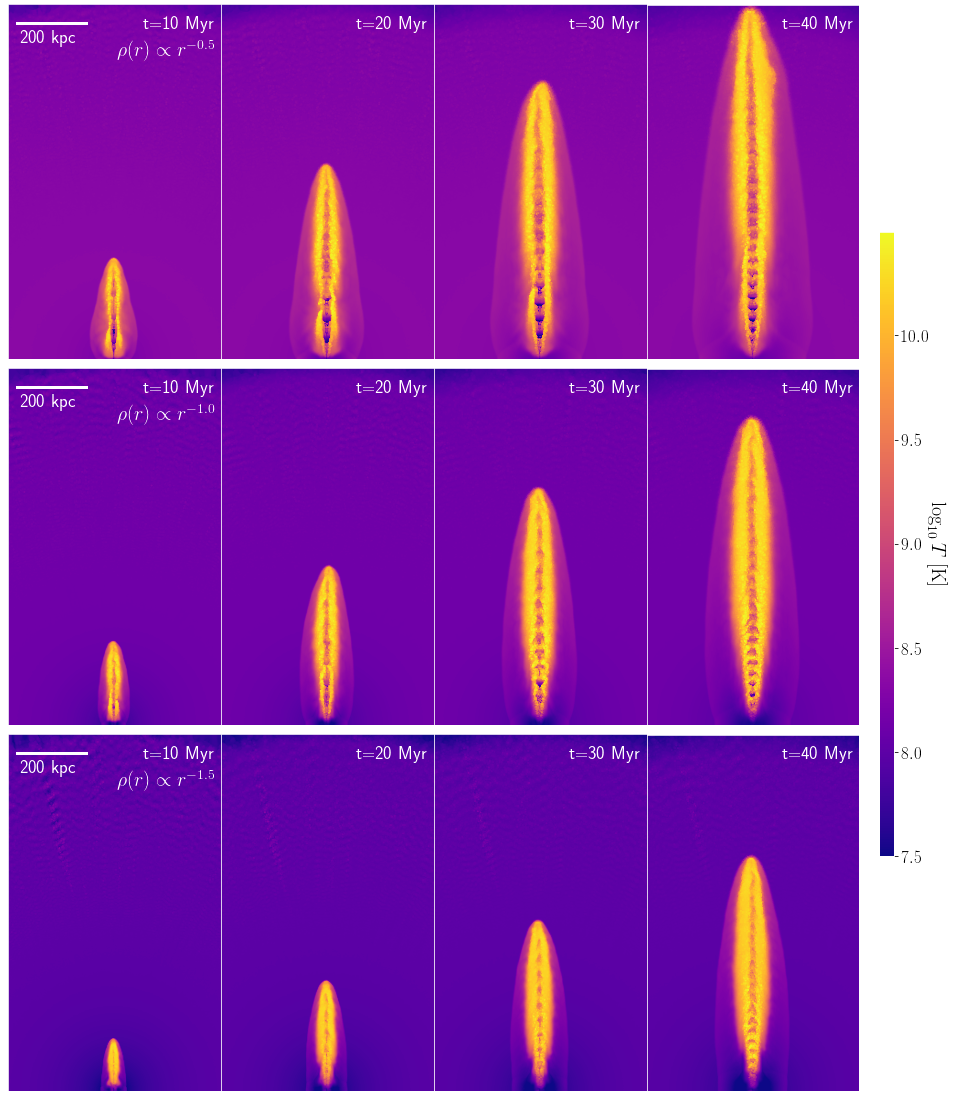}
\caption{Comparison of the evolution of jets launched into power-law ambient density profiles (see top-right-hand corner of the first panel in each row). Other jet-related parameters used are listed in the last row of Table \ref{tab:tab0}. The colours show gas temperature, according to the colour bar. The panels are 25 kpc in depth.}
\label{fig:power_law_vis}
\end{figure*}

In Fig. \ref{fig:power_law_vis} we show a visual comparison of all three power-law simulations we have done, at different times up to $t=40$ Myr. All three simulations feature very similar jets and jet-inflated lobes. However, there are differences between these simulations and the constant-density ones featured in the rest of the paper. Among these is that the jet-inflated lobes appear to have a larger aspect ratio (i.e. they are thinner), despite having the same half-opening angle $\theta_\mr{j}=10\degree$. In addition, the unshocked, colder jet gas appears to extend all the way to the jet head in all three cases. This is a feature also shared by ballistic jets (see top row of Fig. \ref{fig:fig1}). However, these jets also clearly feature multiple recollimation shocks, which only appear in the self-similar regime. Inspecting the unshocked jet gas more closely, one can see that the recollimation shocks appear at distances closer to the origin, while at large distances (close to the jet head), they blend in into a single stream (spine) of unshocked gas, appearing very similar to the unshocked jet gas in ballistic jets (see top panels of Fig. \ref{fig:fig1}).

\begin{figure*}
\includegraphics[width=0.99\textwidth, trim = 0 5 0 0]{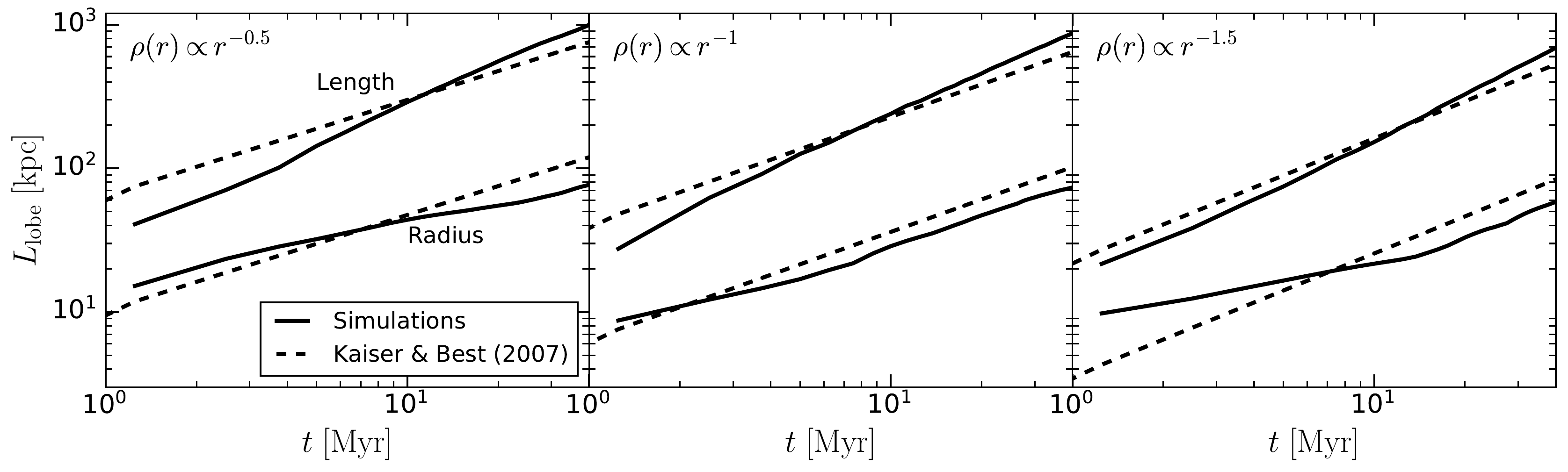}
\caption{Comparison of jet-inflated lobe lengths and radii in simulations with power-law ambient density profiles (see top-left corner of each panel). Other jet-related parameters used are listed in the last row of Table \ref{tab:tab0}. The solid lines show measured lengths and radii from the simulations, while the dashed lines give equivalent predictions for self-similar jet-inflated lobe expansion from \protect\cite{Kaiser2007}.}
\label{fig:power_law_kaiser}
\end{figure*}

In Fig. \ref{fig:power_law_kaiser} we show the time evolution of the lobe lengths and radii for the three simulations compared to predictions from \cite{Kaiser2007}. In all cases, the lobes start off as being shorter and wider than predicted by the self-similar theory. As explained in Section \ref{sec:comparison_sims}, this is potentially a result of the rather long transition from the initial ballistic phase to the self-similar phase. It could also be a result of the initial phase of the jet evolution being relatively poorly resolved (simply due to a small number of particles having been launched into the jets), which tends to make the poorly-resolved lobes more spherical due to SPH averaging effects (see Fig. \ref{fig:figA}). 

At late times, in all three cases the lobes transition to behaviour that is in better agreement with the theoretical predictions. However, the lobes appear to be somewhat too long at late times, while also too thin (both disagreements are at a level of $20-30\%$). These findings confirm the visual appearance of the jets shown in Fig. \ref{fig:power_law_vis}. We posit here that this may be a result of the self-similar theoretical predictions relying on assumptions that may not hold in these simulations. One of these is that the lobe is cylindrical, whereas we find a more ellipsoidal shape. The second assumption is that the lobes have a constant pressure throughout. However, the lobes are more likely to be in pressure equilibrium with the ambient medium locally (i.e. at a given radius). The pressure profile of the ambient medium varies significantly with radius, which means that the lobe has a higher pressure at smaller radii than near the jet head. This could help the jet propagate or the lobes to expand more easily in the radial direction rather than laterally. It could also explain why the recollimation shocks (which occur by means of the lobe pressure acting on the unshocked jet gas) are stronger at smaller radii.

\section{Summary and conclusions}
\label{sec:conclusion}

Using the SWIFT code, and its SPHENIX smoothed particle hydrodynamics scheme, we have simulated the interaction of hydrodynamical jets from active galactic nuclei with the ambient medium up to very high resolutions (more than a million particles per jet, unprecedented for an SPH simulation of jets). These simulations intentionally did not include additional physics, such as gravity, radiative cooling, magnetic fields or cosmic rays, in order for our results to be comparable with simple analytical models of jet evolution.

We launch jets with a constant power and launching velocity, and with finite opening angles, by kicking particles from initially placed reservoirs. These jets propagate through ambient media with constant gas densities as well as power-law density profiles. We find that jets initially evolve ballistically, during which time they drill through the medium with ease, while also launching a bow shock. Once the mass of the swept up ambient medium becomes comparable to that injected into the jet, the jets transition to a self-similar regime. This transition is fairly prolonged, and evidence of the ballistic phase can be seen in the jet morphology even at late times.

The self-similar evolutionary phase is characterised by significant gas shocking, which results in the formation of an ellipsoidal lobe of low-density and high-temperature gas. This shocked gas recollimates and shocks recently launched jet particles. We find that the jet and lobe lengths and lobe radii are in agreement with self-similar predictions where this is expected (for finite half-opening angles and large enough launching velocities). In the self-similar phase, the aspect ratio of the lobes is constant, as is the fraction of initially injected energy contained in the lobes (including the unshocked jets) versus that in the ambient medium. This is also true for the kinetic and thermal components of both the lobes and ambient medium. In our standard constant-density simulation ($m_\mr{gas}=1.81\times10^5$ $\mr{M}_\odot$, $P_\mr{j}=10^{46}$ $\mr{erg}\mr{s}^{-1}$, $v_\mathrm{j}=15000$ km/s and $\theta_\mr{j}=10\degree$), these fractions are: 1) lobe kinetic - $20\%$, 2) lobe thermal - $14\%$, 3) ambient kinetic - $47\%$ and 4) ambient thermal - $19\%$. This shows that energy is effectively transferred from the jet to the ambient gas, although it is not immediately and fully thermalised while the jet is active. We find that these fractions depend on the particular simulation, but none of the components is ever negligible.

We have performed a series of simulations in the constant-density case, with varying parameters such as the jet power, half-opening angle, launching velocity and mass resolution. Increasing the jet power results in longer jets, but also better resolution in the jets due to more particles being launched (at a fixed launching velocity). Increasing the half-opening angle results in shorter jets. We find that jet (lobe) lengths match expectations from analytical models of self-similar evolution for all values of jet power and half-opening angle that we have simulated.

We find that varying the jet launching velocity changes both the physical properties of jets (i.e. temperature and mass of the lobes), as well as the degree to which the jet and lobe are resolved. By varying numerical resolution through the mass of gas particles, we find that our high-resolution ($m_\mr{gas}\approx10^4$ $\mr{M}_\odot$) jet lobes deviate somewhat from the picture expected theoretically. In this case, Kelvin-Helmholtz instabilities prevent the formation of a classical ellipsoidal lobe. For the same reason, the jet spine takes the shape of a conical outflow, and not a series of multiple recollimation shocks, as expected and as found in the lower-resolution cases. We speculate that at even higher resolutions, smaller-scale Kelvin-Helmholtz instabilities would disrupt the large vortices and again give rise to an ellipsoidal lobe. 

We find that high-resolution simulations are not necessary in order for the jets to deposit their energy where they should. On the contrary, we find that very poor resolution ($\approx500$ particles launched per jet) is sufficient for an accuracy of $15$\% (for the lengths and radii of the jet-inflated lobes). We find that the level at which jets are resolved depends not only on numerical resolution directly (through the gas particle mass), but also indirectly on physical parameters such as jet power and launching velocity. This means that different jet episodes will be resolved at different levels depending on the details of the launching scheme and on the energetics of a particular jet episode. We find that resolving a jet with $\approx500$ particles is sufficient to reproduce basic properties such as jet lengths and energetics. The implication of this finding is that jet launching velocities of $\approx10^4$ km/s are appropriate for typical modern cosmological simulations of galaxy clusters. However, we suggest that the jet launching velocity be varied from system to system in order to achieve a balance between resolution and sufficiently strong shock heating of the ambient medium (e.g. through a scaling with the halo mass $v_\mr{j}\propto M_\mr{h}^{\gamma}$, or with the black hole mass $v_\mr{j}\propto M_\mr{BH}^{\gamma}$).

We find that the self-similar regime is also reached in simulations of jets launched into power-law density profiles of the ambient medium ($\rho\propto r^{-\alpha}$, with $\alpha<2$). However, the lobes inflated by these jets appear to be up to $20-30\%$ too long and too thin compared to self-similar predictions. This may be a result of some of the theoretical assumptions breaking down in the power-law case (e.g. pressure equilibrium throughout the lobes), rather than our simulations being inaccurate.

In future, we plan to extend our analysis to more realistic systems (e.g. galaxy cluster in hydrostatic equilibrium) with additional physics included, and to Gyr time-scales (after the jets have been turned off and formed bubbles). We will also study jets as a self-consistent feedback mechanism, where the jet power is calculated based on the spin and environment of a black hole. 




\section*{Acknowledgements}
We thank Josh Borrow for pointing out several errors in the initial manuscript. We also thank the reviewer for a comprehensive report that helped to substantially improve the quality of the paper. The research in this paper made use of the SWIFT open-source simulation code (\url{http://www.swiftsim.com}, \citealt{Schaller2018})
version 0.9.0. The swiftsimio Python library was used to analyze and visualize the data from the simulations (\citealt{Borrow2020_swiftsimio}, \citealt{Borrow_2021_swiftsimio}). This work was supported by the Science Technology Facilities Council through a CDT studentship (ST/P006744/1), and the STFC consolidated grant ST/T000244/1. This work used the DiRAC@Durham facility managed by the Institute for Computational Cosmology on behalf of the STFC DiRAC HPC Facility (www.dirac.ac.uk). The equipment was funded by BEIS capital funding via STFC capital grants ST/K00042X/1, ST/P002293/1, ST/R002371/1 and ST/S002502/1, Durham University and STFC operations grant ST/R000832/1. DiRAC is part of the National e-Infrastructure.

\section*{Data availability}

The data underlying this article are available upon reasonable request to the corresponding author.




\bibliographystyle{mnras}
\bibliography{jet_bibliography} 

\bsp	
\label{lastpage}
\end{document}